\documentclass[preprint,12pt]{elsarticle}

\usepackage{amssymb}
\usepackage{amsmath}
\usepackage{bm}
\usepackage{tikz}

\usepackage{algorithm, algpseudocode}

\usepackage{hyperref}

\usepackage{placeins}

\journal{Computer Physics Communications}

\usepackage[normalem]{ulem}
\usepackage{cancel}
\begin{document}

\begin{frontmatter}



\title{GBEES-GPU: An efficient parallel GPU algorithm for  high-dimensional nonlinear uncertainty propagation}


\author[1,3]{Benjamin L. Hanson} 
\author[2]{Carlos Rubio} 
\author[2]{Adri\'an Garc\'ia-Guti\'errez} 
\author[1]{Thomas Bewley} 

\affiliation[1]{organization={Department of Mechanical and Aerospace Engineering, University of California San Diego},
            addressline={9500 Gilman Dr}, 
            city={La Jolla},
            postcode={92093}, 
            state={CA},
            country={USA}}
\affiliation[2]{organization={Department of Aerospace Engineering, Universidad de León},
            addressline={Av. Facultad de Veterinaria, 25}, 
            city={León},
            postcode={24004},             
            country={Spain}
}
\affiliation[3]{organization = {Corresponding author, Email: blhanson@ucsd.edu, Tel: +1 (913) 626-5877}}

\begin{abstract}

Eulerian nonlinear uncertainty propagation methods often suffer from finite domain limitations and computational inefficiencies. A recent approach to this class of algorithm, Grid-based Bayesian Estimation Exploiting Sparsity, addresses the first challenge by dynamically allocating a discretized grid in regions of phase space where probability is non-negligible. However, the design of the original algorithm causes the second challenge to persist in high-dimensional systems. This paper presents an architectural optimization of the algorithm for CPU implementation, followed by its adaptation to the CUDA framework for single GPU execution. The algorithm is validated for accuracy and convergence, with performance evaluated across distinct GPUs. Tests include propagating a three-dimensional probability distribution subject to the Lorenz '63 model and a six-dimensional probability distribution subject to the Lorenz '96 model. The results imply that the improvements made result in a speedup of over 1000 times compared to the original implementation.

\end{abstract}

\end{frontmatter}

\section{Introduction} \label{sec_intro}
Uncertainty propagation (UP) is the process by which the initial uncertainty of a state is evolved over time. While closely related to the state estimation problem, UP focuses on the estimation of the state uncertainty rather than the state itself, although the latter may often be extracted from the former. Like state estimation, UP consists of two repeated steps known as \textit{prediction} and \textit{correction}. During prediction, state uncertainty is propagated via the dynamical system governing the state's evolution. At correction intervals, state measurements are sourced from a measurement model and combined with the predicted uncertainty. In this paper, high-dimensional UP refers to cases where the state dimension $n>3$, which most commonly arise when the state depends on the position-velocity phase space. This is prevalent in fields such as orbital mechanics and space surveillance \cite{poore_covariance_2016,vetrisano_uncertainty_2016}, attitude estimation \cite{vetrisano_uncertainty_2016}, aircraft navigation \cite{sankararaman_uncertainty_2017}, robotics and robust control \cite{smith_uncertainty_2014}, weather prediction and climate models \cite{smith_uncertainty_2014}, hydrology and geology \cite{smith_uncertainty_2014}, nuclear physics \cite{smith_uncertainty_2014}, biological and chemical models \cite{smith_uncertainty_2014,najm_uncertainty_2009}, reliability and safety studies of structural designs \cite{sudret_uncertainty_2007}, and many others.

When the dynamical system and measurement model are linear, and the initial state uncertainty is Gaussian, state uncertainty remains Gaussian globally; therefore, only estimation of the mean and covariance is required for complete representation of the uncertainty, a tractable problem optimally solved by Kalman \cite{kalman}. The Kalman filter is suboptimal when one or both of the dynamical system and measurement model are nonlinear. In this case, an infinite number of parameters may be required to represent the  uncertainty. Nonlinear uncertainty propagation (NUP) numerical methods aim to represent this intractable uncertainty with finite abstractions. These abstractions must be computed efficiently while preserving accuracy, and typically adhere to one of three methodologies: Kalman, Langragian, and Eulerian.

The Eulerian approach to NUP considers and evolves uncertainty by discretizing phase space on a grid. Godunov \cite{Godunov} and Lax and Wendroff \cite{pL05} designed first-order accurate, grid-based numerical 
schemes using conservation laws, but Arulampalam et al. \cite{arulampalam2002tutorial} highlighted that standard grid-based methods predefined on finite domain spaces suffer from heavy computational cost. Bewley and Sharma \cite{tB12} addressed this limitation with Grid-based Bayesian Estimation Exploiting Sparsity (GBEES), a novel unstructured gridding scheme that excels at NUP when uncertainty is highly non-Gaussian, tracking uncertainty only where it is non-negligible. Outside of the examples discussed in this paper, GBEES has been validated via the NUP of the Poincar\'e orbital element dynamics (2D), a Saturn-Enceladus Distant Prograde Orbit (4D), and a Jupiter-Europa Low Prograde Orbit (6D) \cite{hanson_guide_2025}. A review of NUP methods by Hanson et al. \cite{bH25} validated the accuracy of GBEES but emphasized its inefficiency compared with Kalman and Lagrangian approaches. Davis et al. \cite{davis2025parallel} and Castro et al. \cite{castro2021implementation} employed adaptive mesh refinement paired with grids stored in hash tables in attempts to address this computational cost; because these algorithms are primarily used in computational fluid dynamics (CFD), rarely do they address high-dimensional systems and their complexity.

Of the three numerical methodologies, we assert that the Eulerian approach addresses the problem of NUP most fundamentally, requiring the least abstraction to generally represent and propagate uncertainty. Eulerian approaches do not linearize during prediction or correction, do not require splitting procedures to maintain accuracy, and do not succumb to particle degeneracy. They also happen to be the least explored of the three for high-dimensional systems. Given the primary drawback of this class of methods is computational cost, we argue that the Eulerian approach warrants further investigation for high-dimensional NUP as computational solutions continue to emerge. GPUs are one such promising solution, given that the finite volume schemes are embarrassingly parallelizable, and have been employed often by CFD algorithms for two- and three-dimensional systems (e.g., Ji et al. \cite{ji_gpu-accelerated_2015} Jaber et al. \cite{jaber_gpu-native_2025}). These parallelized algorithms seldom extend to high-dimensional systems.

To address the limitations of high-dimensional Eulerian NUP methods, we introduce GBEES-GPU: a parallel extension of GBEES with improved efficiency achieved by:
\begin{enumerate}
    \item Storing the dynamic grid in a hashtable
    \item Time-marching with a CFL-minimized adaptive step size
    \item Employing directional growing and pruning procedures
    \item Translating the algorithm to CUDA for single GPU execution
\end{enumerate}
These changes result in an efficient, high-dimensional parallel GPU algorithm for NUP, detailed in the remainder of the paper as follows: In Section \ref{sec_gbees} the finite volume formulation underlying GBEES is extended to $n$-dimensions. Section \ref{sec_cpu_adv} outlines the first key deliverable referenced in the abstract, the improvements made to the CPU implementation, while Section \ref{sec_cuda_impl} describes the second key deliverable referenced in the abstract, its adaptation to the CUDA architecture. Section \ref{sec_validation} presents the use cases employed for validation and testing, including a quantitative measure of accuracy and the evaluation of convergence. Next, Section \ref{sec_performance} compares the performance of the improved CPU and GPU implementations against the legacy version. Conclusions are presented in Section \ref{sec_conclusions}, and the hardware specifications for the tests are provided in \ref{appendix_hardware}.

\section{Grid-based Bayesian Estimation Exploiting Sparsity} \label{sec_gbees}

The equations of motion of a stochastic process $\bm{X}(t)\in \mathbb{R}^n$ governed by a combination of deterministic and random forces can be described by the following stochastic differential equation:
\begin{equation}
\label{eq_SDE}
    d\bm{X}(t) = \bm{f}(\bm{X}(t), t)dt + \bm{q}(\bm{X}(t), t)d\bm{W}(t),
\end{equation}
where $\bm{f}(\bm{X}(t), t)$ is the drift vector, $\bm{q}(\bm{X}(t), t)$ is the diffusion vector, and $d\bm{W}(t)=\boldsymbol{\xi}(t)dt$ is a Wiener process, meaning $\boldsymbol{\xi}(t)$ is zero-mean, uncorrelated white noise (i.e., $\mathbb{E}[\boldsymbol{\xi}(t)] = \boldsymbol{0}$ and $\mathbb{E}[\boldsymbol{\xi}(t+\tau)\boldsymbol{\xi}^T(t)] = \boldsymbol{\delta}(\tau)$, where $\mathbb{E}[\cdot]$ is the expectation of $\cdot$). In continuous-time, the Fokker-Planck equation gives the evolution of the probability density function (PDF) $p(\bm{x}, t)$ of $\bm{X}(t)$ in Eq. \eqref{eq_SDE} as follows:
\begin{equation}
    \label{eq_FPE}
    \frac{\partial p(\bm{x}, t)}{\partial t} = -\sum\limits_{j=1}^n\frac{\partial f_j(\bm{x},t)p(\bm{x},t)}{\partial x_j} + \frac{1}{2}\sum\limits_{j=1}^n\sum\limits_{\ell=1}^n\frac{\partial^2 Q_{j\ell}(\bm{x}, t)\,p(\bm{x},t)}{\partial x_j \partial x_{\ell}}
\end{equation}
where $\bm{x} = (x_1, \dots, x_n)$ is a realization of the random variable $\bm{X}(t)$, $f_j(\bm{x}, t)$ is the $j^{\text{th}}$ component of $\bm{f}(\bm{x}, t)$, and $Q_{j\ell}(\bm{x}, t)$ is the $(j, \ell)^{\text{th}}$ component of the process noise spectral density matrix $Q(\bm{x}, t)= \bm{q}(\bm{x},t)\bm{q}^T(\bm{x},t)$. If $Q(\bm{x},t) > 0$ (i.e., it is positive definite), Eq. \eqref{eq_FPE} is elliptic, but if $Q(\bm{x},t)$ is relatively small compared to the deterministic forces, Eq. \eqref{eq_FPE} is hyperbolic and satisfies the conservative form of the $n$-dimensional advection equation: 
\begin{equation}
    \label{eq_conservation_law}
    \frac{\partial p(\bm{x},t)}{\partial t} + \sum\limits_{j=1}^n\frac{\partial  f_j'(\bm{x}, t)}{\partial x_j} = 0,
\end{equation}
where $f_j'(\bm{x}, t)=f_j(\bm{x}, t)p(\bm{x}, t)$. At discrete measurement intervals $t^{(k)}$ the PDF is updated via Bayes' theorem:
\begin{equation}
\label{eq_bayes}
     p(\bm{x},t^{(k+)}) = \frac{p(\bm{y}^{(k)}|\bm{x})\,p(\bm{x},t^{(k-)})}{C},
\end{equation}
where $p(\bm{x},t^{(k+)})$ is the \textit{a posteriori}, $p(\bm{y}^{(k)}|\bm{x})$ is the measurement likelihood, $p(\bm{x},t^{(k-)})$ is the \textit{a priori}, and $C$ is a normalization constant. GBEES performs the accurate, mixed continuous/discrete time-marching of $p(\bm{x},t)$ using numerical approximations of Eqs. \eqref{eq_conservation_law} and \eqref{eq_bayes}. We first delve into the continuous-time prediction of $p(\bm{x},t)$ via a fully discrete flux-differencing method.

\subsection{Corner Transport Upwinding for $n$-dimensional systems} 
\label{sec_flux_differencing}
We use the notation from Colella et al. \cite{colella_2011_high} to define an $n$-dimensional control volume $V_{\bm{i}}$ as 
\begin{equation}
    \label{eq_grid_cell}
    V_{\bm{i}} = \prod_{j=1}^n\Bigg[x_{i_j} - \frac{h_{j}}{2}, x_{i_j} + \frac{h_{j}}{2}\Bigg],
\end{equation}
where the multi-index $\bm{i} = (i_1, \dots, i_n)\in \mathbb{Z}^n$ identifies the control volume center within the grid, and $\bm{h} = (h_1, \dots, h_n) \in \mathbb{R}_+^n$ is the grid spacing vector. From Eq. \eqref{eq_conservation_law}, $p$ is assumed to be conserved over $V_{\bm{i}}$, thus the integral of $p$ varies only due to flux across the boundaries of $V_{\bm{i}}$. The components of the flux vectors at the forward and backward grid cell interfaces of $V_{\bm{i}}$ at time step $t^{(k)}$ are defined as 
\begin{equation}
    \label{eq_flux}
    F^{(k)}_{\bm{i} \pm \frac{1}{2}\bm{e}^j,\,j} \approx \frac{1}{\Delta t \prod\limits_{\substack{{\ell}=1 \\ {\ell}\neq j}}^n {h}_{\ell}} \int_{t^{(k)}}^{t^{(k+1)}} \int_{A_{\bm{i},j}} f_j'(\bm{x}_{\bm{i} \pm \frac{1}{2}\bm{e}^j}, t)d\bm{A}dt.
\end{equation}
where $A_{\bm{i},j}$ are the faces bounding the $V_{\bm{i}}$ with normals pointing in the $j^{\text{th}}$ coordinate direction and $\bm{e}^j$ denotes the unit vector in the $j^{\text{th}}$ coordinate direction. The numerical fluxes $F^{(k)}_{\bm{i} \pm \frac{1}{2}\bm{e}^j,\,j}$ are approximated via a Godunov-type finite volume method known as Corner Transport Upwinding (CTU). We now describe the $n$-dimensional generalization for the CTU method.

First, the Donor Cell Upwind (DCU) method is used to calculate the first-order accurate numerical fluxes. For $\bm{i}\in\mathcal{I}$, where $\mathcal{I}$ represents the complete set of multi-index vectors in the grid, the upwind flux in the $j^{\text{th}}$ direction at time step $k$ is calculated:
\begin{equation}
    \label{eq_DCU}
    F_{\bm{i} - \frac{1}{2}\bm{e}^j, \, j}^{(k)} = f_{\bm{i} - \frac{1}{2}\bm{e}^j, \, j}^+\,P_{\bm{i} - \bm{e}^j}^{(k)} + f_{\bm{i} - \frac{1}{2}\bm{e}^j, \, j}^-\,P_{\bm{i}}^{(k)},
\end{equation}
where $f_{\bm{i} - \frac{1}{2}\bm{e}^j, \, j}^{\pm} = \max/\min(f_{\bm{i} - \frac{1}{2}\bm{e}^j, \, j}^{(k)}, 0)$, $f_{\bm{i} - \frac{1}{2}\bm{e}^j, \, j}^{(k)} = f_j(\bm{x}_{\bm{i}-\frac{1}{2}\bm{e}^j}, t^{(k)})$, and $P^{(k)}_{\bm{i}}$ is the probability defined at grid cell $\bm{i}$ at time step $t^{(k)}$.

Eq. \eqref{eq_DCU} only considers probability flowing normal to the grid cell interface, but in general, probability may flow oblique to the interfaces of $V_{\bm{i}}$. To obtain second-order accuracy, we must account for this with flux corrections. Because the $n$-dimensional CTU method is notationally complex, we provide it in full in Algorithm \ref{alg_CTU}.
\begin{algorithm}[!ht]
\caption{Corner Transport Upwind}\label{alg_CTU}
\begin{algorithmic}[1]
\Require Perform DCU for  $\bm{i}\in\mathcal{I}$ 
\For{$\bm{i}\in\mathcal{I}$}
    \For{$j=1$ to $n$}
        \State $F^* \gets \frac{\Delta t}{2h_j}\big(P_{\bm{i}}^{(k)} - P_{\bm{i}-\bm{e}^j}^{(k)}\big)$
        \For{$\ell = 1$ to $n$, $\ell \neq j$}
            \State $F_{\bm{i} + \frac{1}{2}\bm{e}^{\ell}, \, \ell}^{(k)} \gets F_{\bm{i} + \frac{1}{2}\bm{e}^{\ell}, \, \ell}^{(k)} - f_{\bm{i} - \frac{1}{2}\bm{e}^j,\,j}^+\,f_{\bm{i} + \frac{1}{2}\bm{e}^{\ell},\,\ell}^+\,F^*$
            \State $F_{\bm{i} - \frac{1}{2}\bm{e}^{\ell},\,\ell}^{(k)} \gets F_{\bm{i} - \frac{1}{2}\bm{e}^{\ell},\,\ell}^{(k)} - f_{\bm{i} - \frac{1}{2}\bm{e}^j,\,j}^+\,f_{\bm{i} - \frac{1}{2}\bm{e}^{\ell},\,\ell}^-\,F^*$
            \State $F_{\bm{i} - \bm{e}^j + \frac{1}{2}\bm{e}^{\ell}, \,\ell}^{(k)} \gets F_{\bm{i} - \bm{e}^j + \frac{1}{2}\bm{e}^{\ell}, \,\ell}^{(k)} - f_{\bm{i} - \frac{1}{2}\bm{e}^j, \,j}^-\,f_{\bm{i} - \bm{e}^j + \frac{1}{2}\bm{e}^\ell,\,\ell}^+\,F^*$
            \State $F_{\bm{ i} - \bm{e}^j - \frac{1}{2}\bm{e}^{\ell}, \, \ell}^{(k)} \gets F_{\bm{ i} - \bm{e}^j - \frac{1}{2}\bm{e}^{\ell}, \, \ell}^{(k)} - f_{\bm{i} - \frac{1}{2}\bm{e}^j, \, j}^-\,f_{\bm{i} - \bm{e}^j - \frac{1}{2}\bm{e}^{\ell}, \, \ell}^-\,F^*$
        \EndFor
    \EndFor
\EndFor
\end{algorithmic}
\end{algorithm}

The CTU method is still not second-order accurate, as it is missing high-resolution correction terms. For brevity, and because these correction terms do not depend on dimensionality, we do not restate them here. Instead, they may be found in Bewley and Sharma \cite{tB12}. In totality, Eq. \eqref{eq_DCU}, Algorithm \ref{alg_CTU}, and the high-resolution correction terms result in a Godunov finite volume scheme that is formally second-order accurate; the truncation error analysis proving such may also be found in \cite{tB12}. Given GBEES employs a regular, structured grid scheme, this analysis is sufficient evidence to claim second-order accuracy, as reported by Veluri et al. \cite{veluri2012comprehensive} and Diskin and Thomas \cite{diskin2010notes}.

\begin{figure}[t]
    \centering
    \includegraphics[width=3.0in]{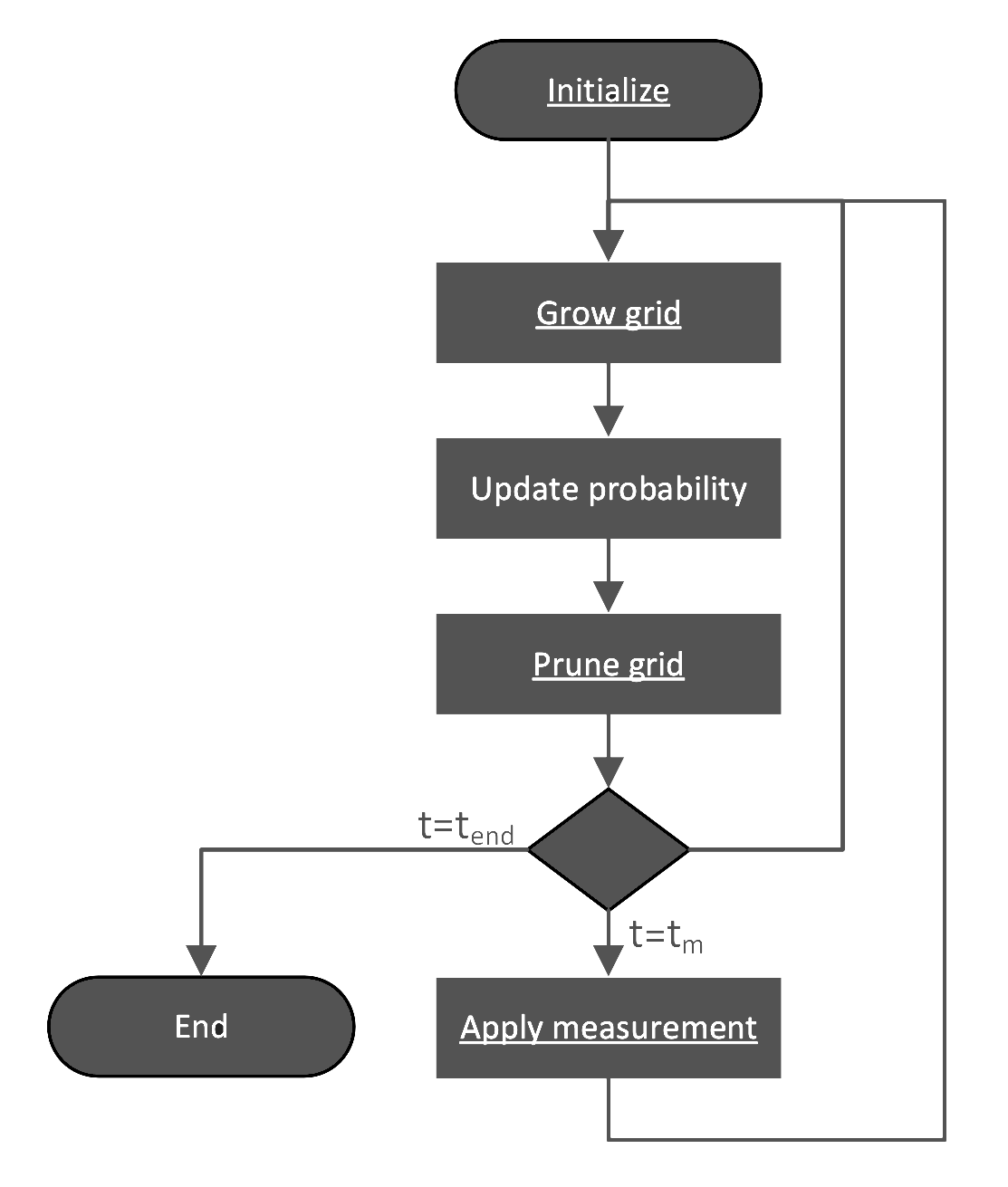}
    \caption{GBEES simplified flowchart. The underscored operations modify the grid by adding or removing active cells. The PDF is propagated until the end time $t_{\text{end}}$ or the time for the next measurement update $t_m$.}
    \label{fig_flowChart}
\end{figure}

\section{Advancements made to the CPU implementation} \label{sec_cpu_adv}
GBEES excels where other finite volume methods fail by dynamically allocating grid cells where probability is above some threshold. Fig. \ref{fig_flowChart} illustrates this process: in the grow grid phase, new cells are added to the discrete representation of the PDF, and in the prune grid phase, cells with probability below the threshold are discarded. This dynamic grid evolution is iterated either until the end time of the uncertainty propagation is reached or until the next measurement update occurs, at which point the grid-based PDF is updated using a Bayesian approach.

This process is implemented in the legacy algorithm; however, that implementation includes structures and subprocesses that are ripe for optimization. Prior to detailing the GPU implementation, we discuss the efficiency improvements made to the CPU implementation.
\subsection{Dynamic grid stored in hashtable} 
\label{sec_hashtable_intro}
The legacy implementation of GBEES stores the dynamic grid in a nested list data structure. Many functions within GBEES require a searching procedure to check if a given $V_{\bm{i}}$ exists in the grid. The time complexity of searching a nested list is $\mathcal{O}(N^2)$, which will result in computational bottlenecks for high-dimensional systems. The first attempt to address this issue employed a binary search tree \cite{bH24}, but overhead of the conversion from grid cell index vector $\bm{i}$ to unique, positive key value proved too large. Instead, a hashtable was utilized, as the structure allows for collisions between mappings, thus removing the overhead from ensuring bijectivity. Additionally, the time complexity of search for a hashtable is $\mathcal{O}(1)$. Hashtables are discussed further in Section \ref{sec_hashtable}. 
\subsection{CFL-minimized adaptive time-marching} \label{sec_cfl_min}
For finite volume methods, the Courant–Friedrichs–Lewy (CFL) condition \cite{CFL} $C\leq C_{\text{max}}$ must be satisfied in order for the method to be stable. The legacy implementation of GBEES employed a static, over-conservative time step to ensure stability for the entire propagation period. The new implementation of GBEES uses the following adaptive time step: 
\begin{equation}
    \label{eq_cfl_3}
    \Delta t^{(k)} = \min\limits_{\bm{i}\in\mathcal{I}} \Bigg[\bigg(\sum_{j=1}^n \frac{f_{\bm{i} - \frac{1}{2}\bm{e}^j,\,j}^{(k)}}{h_j}\bigg)^{-1}\Bigg].
\end{equation}
Implementing Eq. \eqref{eq_cfl_3} maximizes the time step size while ensuring the stability of the explicit finite volume method.

\subsection{Directional growing and pruning} \label{sec_grow_prune}

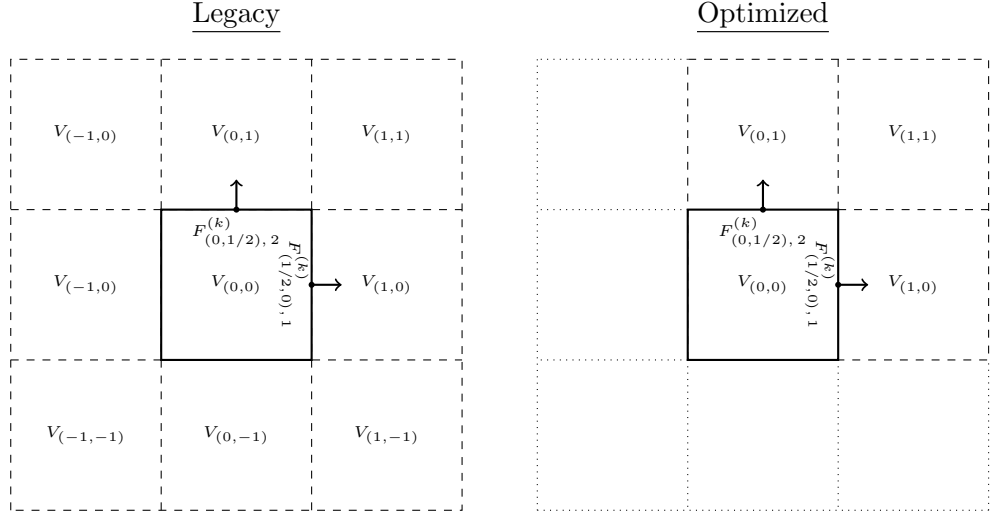
\begin{figure}[t]
    \centering
    \begin{tikzpicture}[scale=2]

    \node[above, font=\small] at (1.5, 3.1) {\underline{Legacy}};
    \node[above, font=\small] at (5, 3.1) {\underline{Optimized}};
    
    \draw[black, thick] (1,1) rectangle (2,2);
    \draw[black, dashed] (1,0) -- (1,3);
    \draw[black, dashed] (2,0) -- (2,3);
    \draw[black, dashed] (0,1) -- (3,1);
    \draw[black, dashed] (0,2) -- (3,2);
    \draw[black, dashed] (0,0) rectangle (3,3);

    \draw[black, thick] (4.5,1) rectangle (5.5,2);
    \draw[black, dashed] (5.5,2) -- (6.5,2);
    \draw[black, dashed] (5.5,2) -- (5.5,3);
    \draw[black, dotted] (4.5,3) -- (3.5,3);
    \draw[black, dotted] (4.5,2) -- (3.5,2);
    \draw[black, dotted] (4.5,1) -- (3.5,1);
    \draw[black, dotted] (4.5,1) -- (4.5,0);
    \draw[black, dotted] (5.5,1) -- (5.5,0);
    \draw[black, dotted] (3.5,3) -- (3.5,0);
    \draw[black, dotted] (3.5,0) -- (6.5,0);
    \draw[black, dotted] (6.5,0) -- (6.5,1);
    \draw[black, dashed] (4.5,1) rectangle (6.5,3);

    \node[anchor=center, font=\tiny] at (0.5,0.5) {$V_{(-1,-1)}$};
    \node[anchor=center, font=\tiny] at (0.5,1.5) {$V_{(-1,0)}$};
    \node[anchor=center, font=\tiny] at (0.5,2.5) {$V_{(-1,0)}$};
    \node[anchor=center, font=\tiny] at (1.5,0.5) {$V_{(0,-1)}$};
    \node[anchor=center, font=\tiny] at (1.5,1.5) {$V_{(0,0)}$};
    \node[anchor=center, font=\tiny] at (1.5,2.5) {$V_{(0,1)}$};
    \node[anchor=center, font=\tiny] at (2.5,0.5) {$V_{(1,-1)}$};
    \node[anchor=center, font=\tiny] at (2.5,1.5) {$V_{(1,0)}$};
    \node[anchor=center, font=\tiny] at (2.5,2.5) {$V_{(1,1)}$};

    \node[anchor=center, font=\tiny] at (5,1.5) {$V_{(0,0)}$};
    \node[anchor=center, font=\tiny] at (5,2.5) {$V_{(0,1)}$};
    \node[anchor=center, font=\tiny] at (6,1.5) {$V_{(1,0)}$};
    \node[anchor=center, font=\tiny] at (6,2.5) {$V_{(1,1)}$};

    \fill[black] (1.5,2) circle (0.02 cm);
    \draw[thick, black, ->] (1.5,2) -- (1.5,2.2);
    \node[anchor=center, font=\tiny] at (1.5,1.85) {$F_{(0, 1/2),\,2}^{(k)}$};
    \fill[black] (2,1.5) circle (0.02 cm);
    \draw[thick, black, ->] (2,1.5) -- (2.2,1.5);
    \node[rotate=270, anchor=center, font=\tiny] at (1.88,1.5) {$F_{(1/2, 0),\,1}^{(k)}$};

    \fill[black] (5,2) circle (0.02 cm);
    \draw[thick, black, ->] (5,2) -- (5,2.2);
    \node[anchor=center, font=\tiny] at (5,1.85) {$F_{(0,1/2),\,2}^{(k)}$};
    \fill[black] (5.5,1.5) circle (0.02 cm);
    \draw[thick, black, ->] (5.5,1.5) -- (5.7,1.5);
    \node[rotate=270, anchor=center, font=\tiny] at (5.38,1.5) {$F_{(1/2, 0),\,1}^{(k)}$};
    
    \end{tikzpicture}
    \caption{2D schematic demonstrating the differences in the growing procedure for the legacy and optimized implementations of GBEES. Solid border cells are those with probability above threshold, dashed border cells are those set to be created during the growing procedure, and dotted border cells are neglected.}
    \label{fig_grow}
\end{figure}

To exploit the sparsity of an $n$-dimensional PDF over phase space, grid cells are tracked where probability is above some threshold $p^*$. To ensure probability is not lost during time-marching, grid cells neighboring those above threshold are also tracked. In the legacy implementation of GBEES, during the growing procedure, the algorithm loops through all existing grid cells and checks if any of the $3^n-1$ neighbors that do not exist must be inserted, regardless if probability is likely to flow into the new grid cell in the following step. This can create irrelevant grid cells that are deleted in future steps without ever increasing in probability. In the new GBEES implementation, the direction of the advection is used to determine if a neighboring grid cell is required for the next time step. As is demonstrated in Fig. \ref{fig_grow}, only the \textit{downwind} grid cells are created; at maximum, this results in checking only $2^n-1$ neighbors, saving on the number of cells that are inserted in each growth step. 

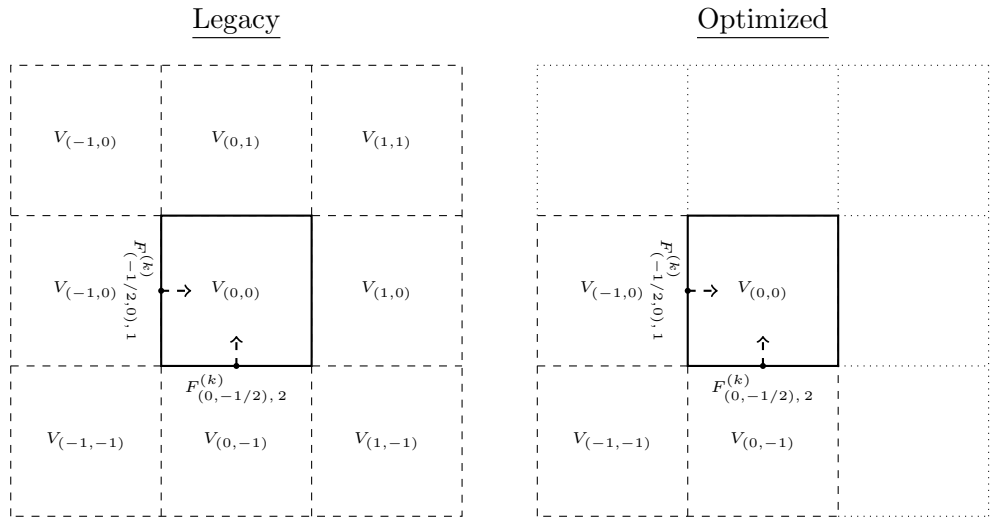
\begin{figure}[t]
    \centering
    \begin{tikzpicture}[scale=2]

    \node[above, font=\small] at (1.5, 3.1) {\underline{Legacy}};
    \node[above, font=\small] at (5, 3.1) {\underline{Optimized}};
    
    \draw[black, thick] (1,1) rectangle (2,2);
    \draw[black, dashed] (1,0) -- (1,3);
    \draw[black, dashed] (2,0) -- (2,3);
    \draw[black, dashed] (0,1) -- (3,1);
    \draw[black, dashed] (0,2) -- (3,2);
    \draw[black, dashed] (0,0) rectangle (3,3);

    \draw[black, thick] (4.5,1) rectangle (5.5,2);
    \draw[black, dashed] (3.5,1) -- (4.5,1);
    \draw[black, dashed] (4.5,0) -- (4.5,1);
    \draw[black, dotted] (3.5,2) -- (3.5,3);
    \draw[black, dotted] (4.5,2) -- (4.5,3);
    \draw[black, dotted] (5.5,2) -- (5.5,3);
    \draw[black, dotted] (6.5,0) -- (6.5,3);
    \draw[black, dotted] (3.5,3) -- (6.5,3);
    \draw[black, dotted] (5.5,2) -- (6.5,2);
    \draw[black, dotted] (5.5,1) -- (6.5,1);
    \draw[black, dotted] (5.5,0) -- (6.5,0);
    \draw[black, dashed] (3.5,0) rectangle (5.5,2);

    \node[anchor=center, font=\tiny] at (0.5,0.5) {$V_{(-1,-1)}$};
    \node[anchor=center, font=\tiny] at (0.5,1.5) {$V_{(-1,0)}$};
    \node[anchor=center, font=\tiny] at (0.5,2.5) {$V_{(-1,0)}$};
    \node[anchor=center, font=\tiny] at (1.5,0.5) {$V_{(0,-1)}$};
    \node[anchor=center, font=\tiny] at (1.5,1.5) {$V_{(0,0)}$};
    \node[anchor=center, font=\tiny] at (1.5,2.5) {$V_{(0,1)}$};
    \node[anchor=center, font=\tiny] at (2.5,0.5) {$V_{(1,-1)}$};
    \node[anchor=center, font=\tiny] at (2.5,1.5) {$V_{(1,0)}$};
    \node[anchor=center, font=\tiny] at (2.5,2.5) {$V_{(1,1)}$};

    \node[anchor=center, font=\tiny] at (5,1.5) {$V_{(0,0)}$};
    \node[anchor=center, font=\tiny] at (5,0.5) {$V_{(0,-1)}$};
    \node[anchor=center, font=\tiny] at (4,1.5) {$V_{(-1,0)}$};
    \node[anchor=center, font=\tiny] at (4,0.5) {$V_{(-1,-1)}$};

    \fill[black] (1.5,1) circle (0.02 cm);
    \draw[thick, dashed, black, ->] (1.5,1) -- (1.5,1.2);
    \node[anchor=center, font=\tiny] at (1.5,0.85) {$F_{(0, -1/2),\,2}^{(k)}$};
    \fill[black] (1,1.5) circle (0.02 cm);
    \draw[thick, dashed, black, ->] (1,1.5) -- (1.2,1.5);
    \node[rotate=270, anchor=center, font=\tiny] at (0.85,1.5) {$F_{(-1/2, 0),\,1}^{(k)}$};

    \fill[black] (5,1) circle (0.02 cm);
    \draw[thick, dashed, black, ->] (5,1) -- (5,1.2);
    \node[anchor=center, font=\tiny] at (5,0.85) {$F_{(0, -1/2),\,2}^{(k)}$};
    \fill[black] (4.5,1.5) circle (0.02 cm);
    \draw[thick, dashed, black, ->] (4.5,1.5) -- (4.7,1.5);
    \node[rotate=270, anchor=center, font=\tiny] at (4.35,1.5) {$F_{(-1/2, 0),\,1}^{(k)}$};
    
    \end{tikzpicture}
    \caption{2D schematic demonstrating the differences in the pruning procedure for the legacy and optimized implementations of GBEES. Solid border cells are those with probability below threshold, dashed border cells are those checked during the pruning procedure, and dotted border cells are neglected.}
    \label{fig_prune}
\end{figure}

Similarly, during the pruning procedure, the algorithm loops through all existing grid cells, looking for those that are below threshold $p^*$. In the legacy GBEES implementation, before deleting the negligible cell, the algorithm checks each of the $3^n-1$ neighbors to see if any are above $p^*$. Again, this results in redundant cells being saved, as even if a neighboring cell is above threshold, it does not necessarily mean that in the following time steps, it will flow probability into the considered cell. Instead, the new GBEES implementation takes a directional-approach to the growth procedure, wherein only the neighboring grid cells that are \textit{upwind} are checked for probability above $p^*$. Fig. \ref{fig_prune} shows that this requires the algorithm to check at maximum, $2^n-1$ neighbors, a fraction of the total neighbor cells, while ensuring that negligible cells are not saved, again contribution to the efficiency of the new algorithm. 

\section{CUDA Implementation} \label{sec_cuda_impl}

The CUDA implementation builds upon the enhancements made to the CPU version described in the previous section. This section details the implementation for the CUDA architecture \cite{nvidia-corporation_about_2024} and the optimization strategies employed.

As described in Section \ref{sec_gbees}, GBEES requires a dynamic grid in which cells are added or removed throughout the integration steps. The scheme depicted in the simplified flowchart of Fig. \ref{fig_flowChart} is representative of the high-level operations of both the CPU and GPU versions. The underscored operations modify the grid by adding or removing active cells. Consequently, the dynamic grid not only changes at each integration step but may also be updated multiple times within a single iteration.

This dynamic nature of the grid prevents establishing a fixed mapping between execution threads and cells. In parallel implementations of finite volume software using a static grid \cite{tang_review_2021,karzhaubayev_dugks-gpu_2024,xue_improved_2021,ye_accelerating_2022}, the grid is partitioned into subdomains (each one including also a boundary with additional halo or ghost cells) and these subdomains are then assigned to thread blocks on the GPU. However, with a dynamic grid, a flexible assignment of cells to threads is required, along with additional synchronization, as thread blocks can no longer operate independently within isolated subdomains.

The problem of handling a dynamic grid alongside GPU parallelization has not been previously addressed in the field of uncertainty propagation. However, a closely related problem arises in Adaptive Mesh Refinement (AMR). AMR \cite{berger_adaptive_1984,berger_local_1989} is used in CFD to refine mesh regions with high gradients, such as shocks or vortices, and is also applied in fields like astrophysics and structural analysis.

AMR involves mesh refinement and coarsening phases, similar to the growth and pruning of the grid in GBEES. Given the difficulty and performance penalty associated with synchronizing the dynamic memory structures on the GPU, some AMR implementations \cite{wang_adaptive_2010,beckingsale_resident_2015,schive_gamer_2010} adopt a hybrid CPU-GPU approach, where mesh modifications occur on the CPU while flow propagation is computed on the GPU. Other studies have pursued a fully GPU-based implementation by leveraging specialized low-level synchronization mechanisms and data structures, including lists/trees \cite{luo_gpu_2016, ji_gpu-accelerated_2015,giuliani_adaptive_2019,jaber_gpu-native_2025} and hashtables \cite{raateland_dcgrid_2022}.

The implementation proposed in this paper combines specific data structures, synchronization algorithms, and parallel techniques to optimize the GBEES-GPU solver. While some elements resemble those used in AMR (such as using hashtables \cite{raateland_dcgrid_2022}, stream compaction \cite{giuliani_adaptive_2019}, and additional lists \cite{luo_gpu_2016}) they are adapted to suit the distinct nature of the problem.

The dynamic structure of the grid impacts all aspects of the GPU implementation, necessitating more complex synchronization mechanisms and a highly efficient memory layout for storing the grid. This extra global-level synchronization required is provided by the Cooperative Kernel abstraction in CUDA \citep{nvidia-corporation_cuda_2024}. A Cooperative Kernel requires all threads to be active concurrently, enabling the establishment of global synchronization barriers. This means the maximum number of threads equals the GPU device's maximum simultaneous threads. If the grid contains more cells than this limit, each thread must process multiple cells sequentially.

The thread-cell assignment is therefore dynamic and varies several times at each integration step. Given a set of $\ell$ threads $(th_1,\dots,th_{\ell})$ and $m$ active cells in the current integration step $(c_1,\dots,c_m)$, each thread $j$ processes sequentially the cells $(c_j, c_{j+\ell}, c_{j+2\ell}, \dots)$. The $\ell$ threads run in parallel, so in the first stage, they process $(c_1, \dots ,c_{\ell})$; in the second stage, they process $(c_{\ell+1}, \dots ,c_{2\ell})$; in the third, $(c_{2\ell+1}, \dots ,c_{3\ell})$; and so on until all active cells in the grid have been processed.

\subsection{Grid data structure} \label{sec_data_struct}

\begin{figure}[!t]
\centering
\includegraphics[width=5.0in]{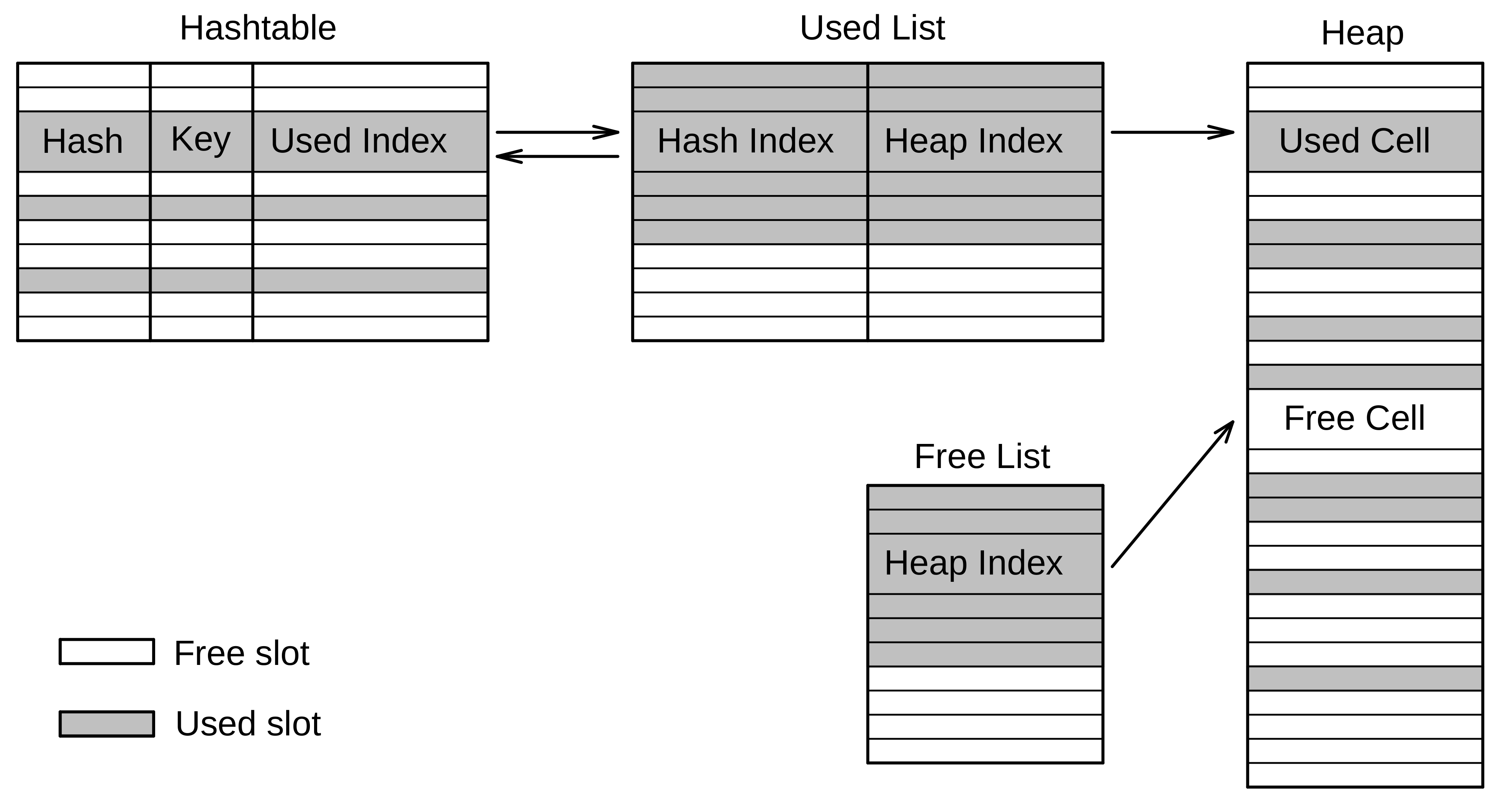}
\caption{Grid data structure. Notice that the Used List and the Free List are maintained compact, with all the used slots at the beginning and all the free slots at the end.}\label{fig_data_structure}
\end{figure}

Given the synchronization requirements, the dynamic thread-cell assignments, and the need for each thread to process a variable number of cells, we propose the data structure for the grid depicted in Fig. \ref{fig_data_structure}. This data structure consists of a hashtable, a list to track used cells, another list for unused cells, and a heap table for cell storage.

The hashtable provides fast access to a cell by its key, the multi-index $\bm{i}=(i_1,\dots,i_n)$. This random access is crucial when accessing the neighbor cells in each dimension. The list of used cells in the grid serves two main purposes. First, it enables quick access to used cells for functions that process individual cells. More importantly, it allows an even distribution of workload among active threads. The Free List maintains a record of unused cells in the heap, reducing the time needed to locate a free slot. Finally, the heap stores the cells themselves. In CUDA, it is not possible to allocate memory directly within a device function in a kernel. Therefore, and for performance reasons, all memory structures are of fixed size, which, for a given configuration, sets the maximum number of cells in the grid.

The relationships among these memory structures are depicted in Fig. \ref{fig_data_structure}. In addition to the depicted relationships, each cell contains pointers to its neighboring cells in each dimension by directly storing their indexes of the Used List.

\subsection{Hashtable} \label{sec_hashtable}

Because the maximum number of grid cells is fixed,  we can set the hashtable size to ensure a maximum occupancy level. The hashtable size is configurable in the software as a multiple of the grid's maximum size, with a default setting of twice that size. This default configuration ensures a maximum occupancy factor $\alpha=0.5$. This bounded occupancy allows us to use a simple open-addressing scheme with linear probing \cite{mehta_handbook_2005}. With a well-randomized hash function \cite{knuth_art_1973}

\begin{equation*}
    \mathbb{E}[\text{\# of probes}] = \begin{cases}
        (1+1/(1-\alpha)^2)/2 & \text{unsuccessful search} \\
        (1+1/(1-\alpha))/2 & \text{successful search}
    \end{cases}.
\end{equation*}

The open addressing scheme requires marking elements as deleted \cite{mehta_handbook_2005}. In GBEES, all cell deletions occur during the prune grid operation. Since the prune operation is executed only for a subset of integration steps, rehashing the hashtable after each grid prune has minimal impact on performance while ensuring that the maximum occupancy level is maintained. Moreover, in the CUDA implementation, this rehashing is fully parallelized, with all execution threads sharing the workload to rehash the hashtable entries concurrently.

Finally, achieving the expected efficiency requires a well-randomized hash function. The key being hashed is the multi-index $\bm{i}$, also known as $n$-gram. Hashing by cyclic polynomial, also known as BuzHash, is an effective method for hashing such $n$-grams, as described in \cite{BuzHash}. Thus, the BuzHash is used by GBEES.

\subsection{Main code blocks} \label{sec_operations}

From an implementation perspective, we can divide the main code blocks into operations that act on individual cells and those that act on the grid as a whole. The first group is straightforward to implement, as each thread modifies only its assigned cells without significant synchronization issues, requiring only quick access to the cell and its neighbors. This rapid access is achieved through the Used List. This category includes operations such as cell initialization, updating references to the neighbor cells, updating the time step based on the CFL condition [Eq. \eqref{eq_cfl_3}], computing the DCU and CTU [Eq. \eqref{eq_DCU}, Algorithm \eqref{alg_CTU}, and high-resolution correction terms], probability distribution normalization, and applying new measurements [Eq. \eqref{eq_bayes}].

The second category involves grid-wide operations, specifically the grid growth and grid pruning. These operations modifies a shared global resource, the grid, and therefore require careful synchronization. To optimize CUDA performance, all synchronization is managed using atomic operations and synchronization barriers, either at the block or device level. The following sections detail the key synchronization aspects and parallel techniques applied in these code blocks. 

\subsection{Synchronization aspects}

\subsubsection{Grow grid operation}
To maximize efficiency in the grid growth operation, a concurrent cell insertion method is required to avoid blocking threads during simultaneous cell creations. Additionally, when exploring different dimensions in the phase space, the algorithm frequently attempts to create cells that already exist. Algorithm \ref{alg_insertCell} presents the chosen compromise solution, which ensures correct synchronization without thread blocking by utilizing atomic operations.

This implementation delays the complete initialization of the cell (using a callback function) until it is confirmed that the cell does not already exist in the grid, thereby improving performance. However, the selected approach has a trade-off: it can only check the existence of a new cell against the previous state of the grid and cannot guarantee successful checking with the other concurrent insertions. To address this limitation, the grid growth operation adopts a staged, directional cell growth strategy:
\begin{enumerate}
    \item Growth is performed along the forward axis of all dimensions. A global synchronization barrier is then executed
    \item Growth is performed along the backward axis of all dimensions, followed by another global synchronization step.
    \item Edge growth is carried out in a similar staged manner in the four diagonal directions (forward-forward, forward-backward, backward-forward, and backward-backward).
\end{enumerate}
This staged approach ensures that no concurrent thread attempts to insert the same cell at the same time.

\begin{algorithm}[t]
\caption{Concurrent Cell Creation}\label{alg_insertCell}
\begin{algorithmic}[1]
\Require usedList and freeList are compact
\State $\text{hash} \gets \text{BuzHash}(\bm{i})$
\For{count $\in$ size(hashtable)}
\Comment{\textbf{linear probing}}
\State $\text{hashIndex} \gets (\text{hash} + \text{count}) \, \% \, \text{size}(\text{hashtable})$
  \If{hashtable[hashIndex] \textbf{is} free}
  \Comment{\textbf{current slot is empty}}
  \State $\text{usedIndex} \gets \text{atomicAdd}[\text{size}(\text{usedList})]$
  \Comment{\textbf{reserve used slot}}
  \State $\text{freeIndex} \gets \text{atomicDec}[\text{size}(\text{freeList})]$
  \Comment{\textbf{reserve free slot}}
  \State $\text{hashtable[hashIndex]}.\bm{i} \gets \bm{i}$
  \Comment{\textbf{update hashtable and lists}}
  \State $\text{usedList}[\text{usedIndex}].\text{heapIndex}\gets \text{freeList}[\text{freeIndex}]$
  \State $\text{usedList}[\text{usedIndex}].\text{hashIndex}\gets \text{hashIndex}$
  \State complete cell initialization with callback function
  \Else 
  \If{$\text{hashtable[hashIndex]}.\bm{i} \textbf{ is } \bm{i}$}
  \Comment{\textbf{cell already exists}}
  \State{break}
  \EndIf
  \EndIf    
\EndFor
\end{algorithmic}
\end{algorithm}

\subsubsection{Prune grid operation}
The prune grid operation, outlined in Algorithm \ref{alg_pruneGrid}, is less performance-critical as it is executed only once every several integration steps. However, it requires specialized techniques to be performed in parallel by all threads. The operation begins by marking cells whose probability values fall below a specified threshold that do not neighbor any cells exceeding this threshold, identifying them as candidates for pruning. The next step involves performing a parallel prefix sum operation \cite{ladner_parallel_1980, hwu_programming_2023} to compact the Used List. The prefix sum, also known as scan, computes cumulative sums over a list to facilitate parallel data compaction. This scan is carried out by the active threads, as detailed in the following section on specific parallel techniques. After the scan, the Used List is compacted using a double-buffer scheme, and the freed slots are added to the Free List via atomic operations. Finally, the Hashtable is rehashed, also employing a double-buffer scheme and distributing the rehashing workload across all active threads.

\begin{algorithm}[!ht]
\caption{Grid Prune Operation}\label{alg_pruneGrid}
\begin{algorithmic}[1]
\For{$\bm{i} \in \mathcal{I}$}
\If{$\bm{i}.p < p^* \textbf{ and } \bm{i} \text{ is not a neighbor}$}
\State $\bm{i} \gets \text{negligible}$
\EndIf
\EndFor
\State perform a prefix sum process of usedList in shared memory
\State complete the prefix sum of usedList in global memory
\State compact usedList and update freeList
\State rehash hashtable
\Ensure{perform a global synchronization at the end of each step.}
\end{algorithmic}
\end{algorithm}

\subsection{Specific parallel techniques} \label{sec_specific}
The GBEES-GPU implementation employs two high-level parallel techniques: parallel reduction and parallel scan. Parallel reduction is utilized to compute the sum of grid cell probabilities for normalizing the distribution. Parallel scan is applied during the prune operation to compact the Used List. These techniques are widely recognized as standard methods \cite{hwu_programming_2023}, and only a brief description is provided here, focusing on their adaptation to the GBEES kernel's context, which involves multiple concurrent blocks and threads processing several cells each.

In the case of parallel reduction, each thread begins by summing the probability value of all the cells assigned to it. Next, a parallel reduction is performed within each thread block, utilizing shared memory. This intra-block reduction employs a sequential addressing scheme to obtain an optimal shared memory access. Once the reduction within shared memory is complete, a global reduction is performed, involving the first thread of each block. Unlike the intra-block reduction, which uses thread synchronization, the outer reduction relies on global barriers. The final result of the reduction is the sum of the probabilities of all cells.

For the scan operation required to compact the Used List, the process begins with a per-block scan using a double buffer in shared memory. Specifically, an inclusive scan with sequential addressing is employed. Unlike parallel reduction, it is not possible to pre-accumulate the values of all cells processed by each thread. Instead, multiple intra-block scans are performed within each block, with the sums orderly accumulated into a global array. Following this, a second outer scan is conducted at the global level by the first thread of each block. Once the corresponding prefix sums are obtained, each thread populates the compacted Used List in parallel and updates the Free List to account for unused or deleted cells.

\section{Validation} \label{sec_validation}

\subsection{Use cases}

In order to validate the GBEES implementation we propagate uncertainty in the Lorenz '63 model (three-dimensional) and the Lorenz '96 model (six-dimensional).

\subsubsection{Lorenz '63}
\label{sec_lorenz_63}
\begin{figure}[!t]
\centering
\includegraphics[width=5in]{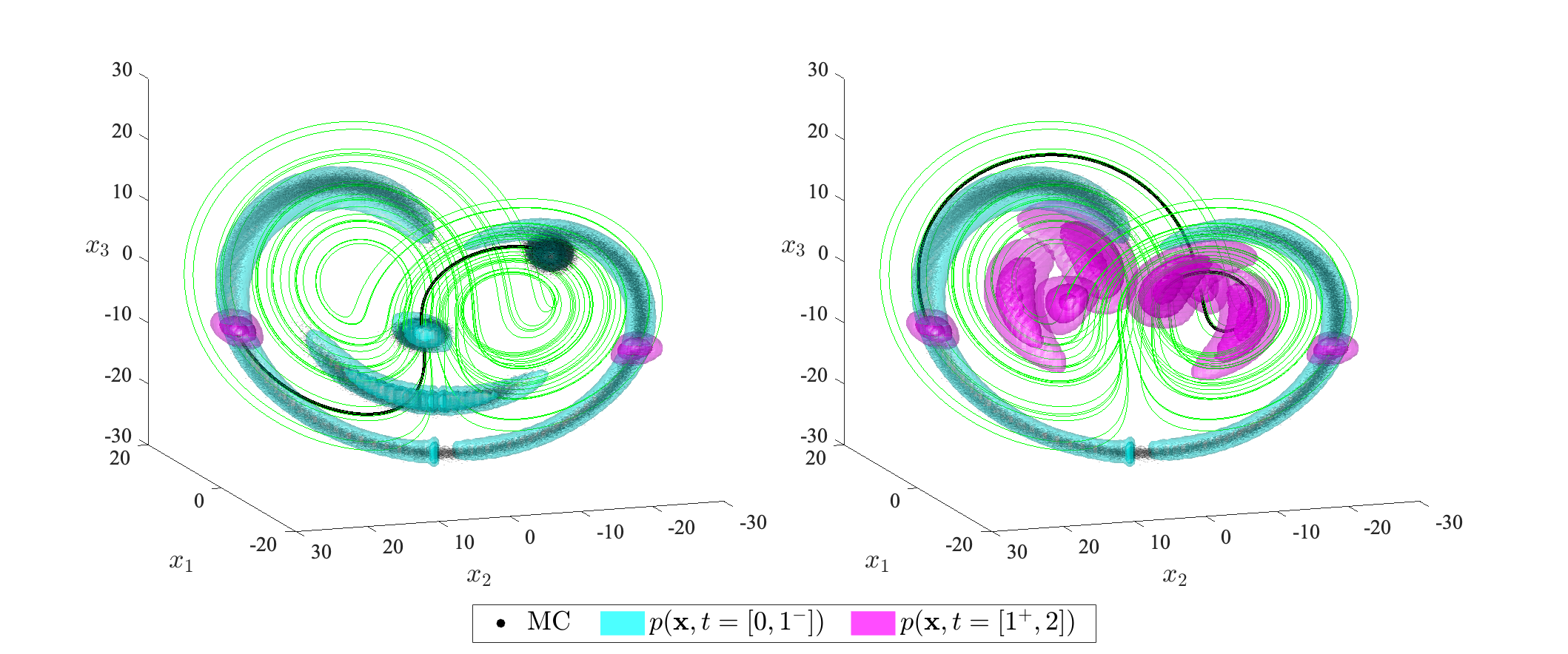}
\caption{PDF isosurfaces governed by the Lorenz '63 model in $(x_1, x_2, x_3)$-space at $p = 0.607$,  $p = 0.135$, and $p = 0.011$ with grid cell width $h_j^* = 0.5$ for $j=1,2,3$, compared with a MC simulation with 100,000 samples. On the left (a), the isosurfaces and MC distributions are at $t=0$, $t=1/3$, $t=2/3$,and $t=1^{\pm}$ and on the right (b), the isosurfaces are at $t=1^{\pm}$, $t=4/3$, $t=5/3$ and $t=2$.}\label{fig_validation_1}
\end{figure}

The Lorenz '63 model, colloquially referred to as the Butterfly Effect, is often employed to validate the accuracy of uncertainty propagation methods because of the highly non-Gaussian behavior exhibited \cite{Lorenz63}. The three-dimensional state and equations of motion are defined as
\begin{equation*}
   \bm{x} = \begin{bmatrix}
       x_1 \\ x_2 \\ x_3
   \end{bmatrix}, \quad \frac{d\bm{x}}{dt} = \bm{f}(\bm{x}) = \begin{bmatrix}
       \sigma (x_2 - x_1) \\ -x_2-x_1x_3 \\ -bx_3+x_1x_2-br
   \end{bmatrix}, 
\end{equation*}
where $(\sigma, b, r) = (4, 1, 48)$ results in the system being chaotic. In Fig. \ref{fig_validation_1}, a 3D Gaussian PDF is initialized at $\bm{x}^{(0)} = [-11.5, -10, 9.5]^T$ with standard deviation $\sigma_{x_j}=1$ for  $j=1,2,3$. The uncertainty is then propagated via the system dynamics until the next measurement correction at $t=1$. To validate the accuracy of GBEES, a Monte Carlo (MC) simulation with identical initial conditions is propagated up to this epoch. Because MC cannot assimilate measurement corrections, the simulation and comparison with GBEES ends here. At $t=1$, a discrete measurement update is performed with measurement $y^{(1)}=-8$, where the measurement model is 
\begin{equation*}
    y = h(\bm{x}) = x_3,
\end{equation*}
and the measurement uncertainty $\sigma_y = 1$. The uncertainty is then propagated via the system dynamics until $t=2$, when the GBEES simulation ends. Fig. \ref{fig_validation_1} illustrates the rapid evolution of the PDF from Gaussian to highly non-Gaussian, with the PDF naturally bifurcating at $t=1$. Since this model was also used to  validate the legacy GBEES implementation, it serves as a valuable basis for performance comparison, as discussed further in Section \ref{sec_performance}.

\begin{figure}[!t]
\centering
\includegraphics[width=5in]{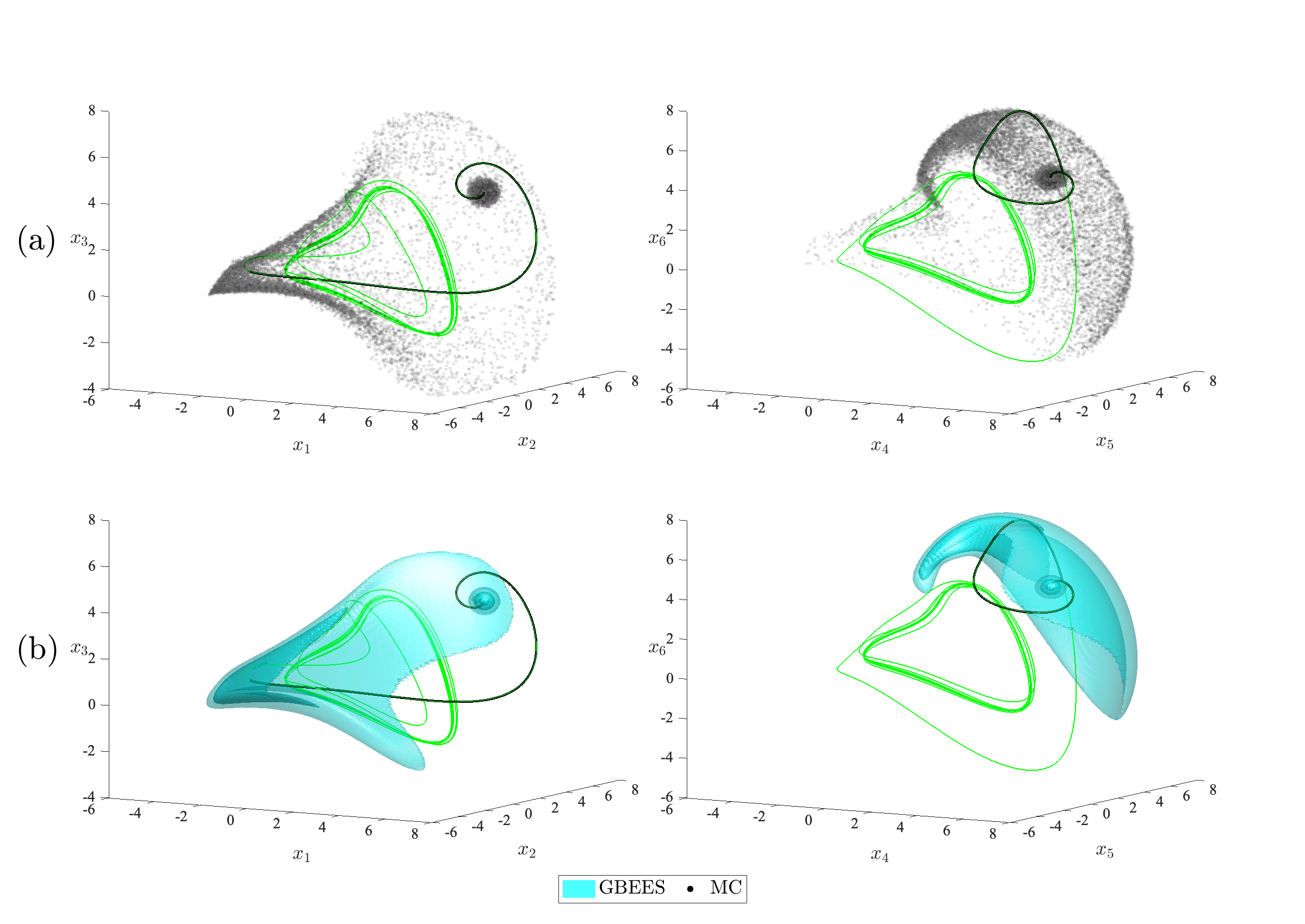}
\caption{PDF isosurfaces in $(x_1, x_2, x_3)$- and $(x_4, x_5, x_6)$-spaces governed by the Lorenz '96 model with grid cell width $h_j^* = 0.1$ for $j=1,\dots,6$ and $F=4$. On top (a), the MC point-mass distributions with 10,000 samples at $t=0$ and $t=1.3$ and on bottom (b), the GBEES isosurfaces at $p = 0.607$,  $p = 0.135$, and $p = 0.011$ for $t=0$ and $t=1.3$.}\label{fig_validation_2}
\end{figure}

\subsubsection{Lorenz '96}
\label{sec_lorenz_96}
As an analog to the Lorenz '63 validation first performed in \cite{tB12}, the CPU-optimized and GPU implementations of GBEES are validated on the Lorenz '96 model, a generalized dynamical system that exhibits chaotic behavior \cite{Lorenz96}. The $n$-dimensional state and equations of motion are defined as
\begin{equation*}
   \bm{x} = \begin{bmatrix}
       x_1 \\ \vdots \\ x_j \\ \vdots \\ x_n
   \end{bmatrix}, \quad \frac{d\bm{x}}{dt} = \bm{f}(\bm{x}) = \begin{bmatrix}
        (x_2 - x_{n-1})x_{n} - x_1 + F \\ \vdots \\ (x_{j+1} - x_{j-2})x_{j-1} - x_j + F \\ \vdots \\ (x_{1} - x_{n-2})x_{n-1} - x_n + F
   \end{bmatrix}, 
\end{equation*}
where $(F, \dots, F)$ is an unstable equilibrium, with $F$ being a forcing constant. A 6D Gaussian PDF is initialized at $\bm{x}^{(0)} = [F + 0.5, F, \dots, F]^T$ where $F=4$, with standard deviation $\sigma_{x_j} = 0.2$ for $j=1,\dots,6$. The uncertainty is then propagated via the system dynamics until $t=1.3$. No measurement update is performed in this simulation. To validate the accuracy of GBEES, an MC simulation with identical initial conditions is plotted, depicted in Fig. \ref{fig_validation_2}(a). The 3D PDFs in the $(x_1, x_2, x_3)$- and $(x_4, x_5, x_6)$-spaces, shown in Fig. \ref{fig_validation_2}(b), are calculated from the discretized 6D PDF by numerically integrating over the $(x_4, x_5, x_6)$- and $(x_1, x_2, x_3)$-spaces, respectively:
\begin{align*}\label{eq_integrate}
    p(x_1,x_2,x_3, t) &= \int_{\min(x_6)}^{\max(x_6)}\int_{\min(x_5)}^{\max(x_5)}\int_{\min(x_4)}^{\max(x_4)} p(\bm{x}, t)dx_4dx_5dx_6, \\ 
    p(x_4, x_5, x_6, t) &=\int_{\min(x_3)}^{\max(x_3)}\int_{\min(x_2)}^{\max(x_2)}\int_{\min(x_1)}^{\max(x_1)} p(\bm{x}, t)dx_1dx_2dx_3.
\end{align*}

\subsection{Accuracy and convergence}

In addition to the qualitative validation against the MC simulation shown in Figs. \ref{fig_validation_1} and \ref{fig_validation_2}, two quantitative validations are performed. The first compares the PDFs obtained using GBEES with those from a dense MC simulation to evaluate the accuracy of GBEES uncertainty propagation. The second  compares different GBEES versions to validate the CUDA implementation and verify proper convergence as the step size is reduced.

\subsubsection{Comparison with MC}

The quantitative comparison with a dense MC simulation is performed using the Bhattacharyya Coefficient (BC) \cite{bhattacharyya_measure_1960}, defined as
\begin{equation}\label{eq_bc}
{BC}(t) = \sum_{\bm{i}\in\mathcal{I} }\sqrt{p(\bm{x}_{\bm{i}},t)p^0(\bm{x}_{\bm{i}},t)}.
\end{equation}
$BC$ measures the similarity between two probability distributions, where $BC=1$ indicates perfect coincidence and $BC=0$ indicates complete dissimilarity. The BC was chosen over other similarity metrics due to $0 \leq BC \leq 1$. A more commonly-used metric, the Kullback–Leibler (KL) divergence \cite{KL}, approaches $\infty$ as $P$ and $Q$ approach complete dissimilarity. To ensure the readability of figures illustrating the time evolution of the similarity between the truth and approximate distributions, the BC was selected for use in this paper.

The computation of $BC$ requires that the MC set of samples be first represented as a PDF function. This representation was obtained using kernel density estimation (KDE) with Gaussian kernels and a bandwidth selection using the Scott factor \cite{scott_multivariate_2015}.
\begin{table}[t!]
\centering
\begin{tabular}{l r r}
\hline
 & Lorenz '63 at $t=1$ & Lorenz '96 at $t=1.3$ \\
\hline
CPU-legacy & 0.9047 &  \\
CPU-optimized & 0.9027 & 0.9155 \\
GPU & 0.9027 & 0.9155 \\
\hline
\end{tabular}
\caption{$BC$ values comparing the Lorenz '63 and Lorenz '96 GBEES propagations with a KDE representation of the Monte Carlo simulation of 100,000 samples. Lorenz '96 was not tested in the CPU-legacy version due to its high computational cost.}
\label{table_MC_validation}
\end{table}
Table \ref{table_MC_validation} shows the $BC$ values for the different GBEES implementations. In all cases, are obtained high values $(>0.9)$, that confirms the similarly observed in Figs. \ref{fig_validation_1} and \ref{fig_validation_2} between the PDFs obtained by the GBEES and the MC method. The specific threshold for sufficient UP accuracy is application-dependent. In this comparison, it is influenced by the number of MC samples, the grid and step sizes of the GBEES configurations, the KDE conversion, and the characteristics of the problem itself.

Finally, note the equality in values between the CPU-optimized and GPU implementations, as both follow the same GBEES algorithm with identical rules for variable step size based on the CFL condition and the same directional growth. That is, the only differences between the CPU-optimized and GPU versions stem from implementation details, which are minimal, as evaluated in the next section.

\subsubsection{Convergence of GBEES}

The second quantitative validation follows the framework for ensuring convergence delineated by Leveque \cite{rL02}. To quantify the global error between an approximate distribution and a truth distribution, for methods that depend on conservation laws, the 1-norm is often used as a convergence metric \cite{medi2011application, schwaiger2012ash3d}, defined as:
\begin{equation*}
    E(t) = \left(\prod_{j=1}^n h_j\right)\sum_{\bm{i}\in\mathcal{I}}\left|p(\bm{x}_{\bm{i}}, t) - p^0(\bm{x}_{\bm{i}}, t)\right|,
\end{equation*}
where $p^0(\bm{x}_{\bm{i}},t)$ represents the truth distribution. A method is then convergent at time $T$ if 
\begin{equation}
    \label{eq_convergence}
    \lim_{\Delta t \rightarrow 0} E(T) = 0.
\end{equation}
Because the CPU-optimized and GPU versions of GBEES use an adaptive step size, we define $\Delta t$ for them as a function of the CFL condition, or $\Delta t = \epsilon\times $CFL. As an analytical truth distribution may not exist in general, we represent $p^0(\bm{x}_{\bm{i}},t)$ using a reference propagation with an extremely small $\epsilon$. As a supplementary metric of convergence, similarly used by Chen et al. \cite{chen2020data}, we also assess the approximate distributions using the BC, as previously defined in Eq. \eqref{eq_bc}.

\begin{figure}[t]
\centering
\includegraphics[width=5.4in]{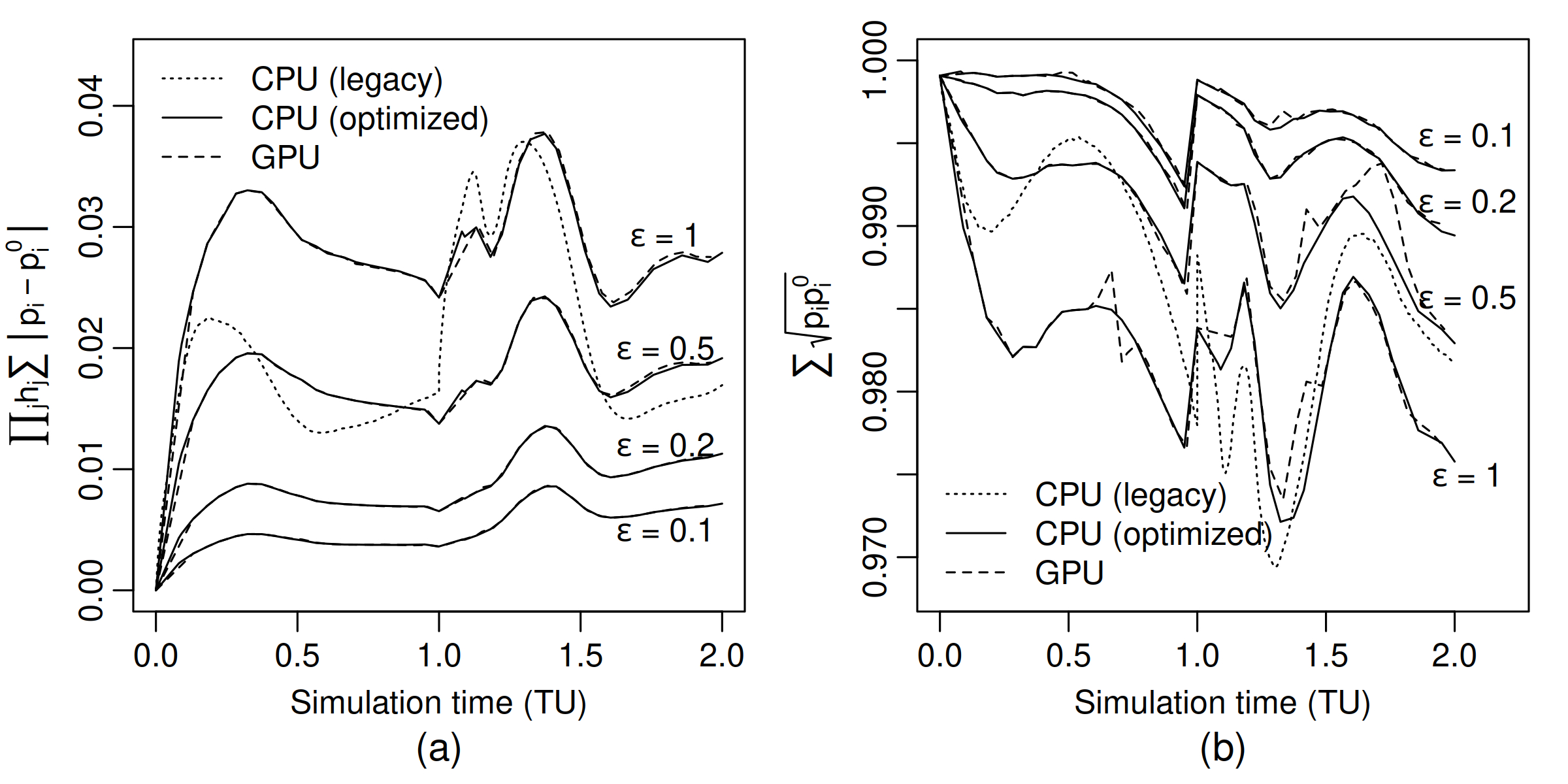}
\caption{Lorenz '63 comparison of the discrete probability distributions for different time step sizes (with respect to a reference propagation with $\epsilon = 0.01$). On the left (a), the comparison is based on the 1-norm $E(t)$. On the right (b), the comparison utilizes the Bhattacharyya Coefficient ${BC}(t)$.}\label{fig_validationMetrics}
\end{figure}

\begin{figure}[t]
    \centering
    \includegraphics[width=5.4in]{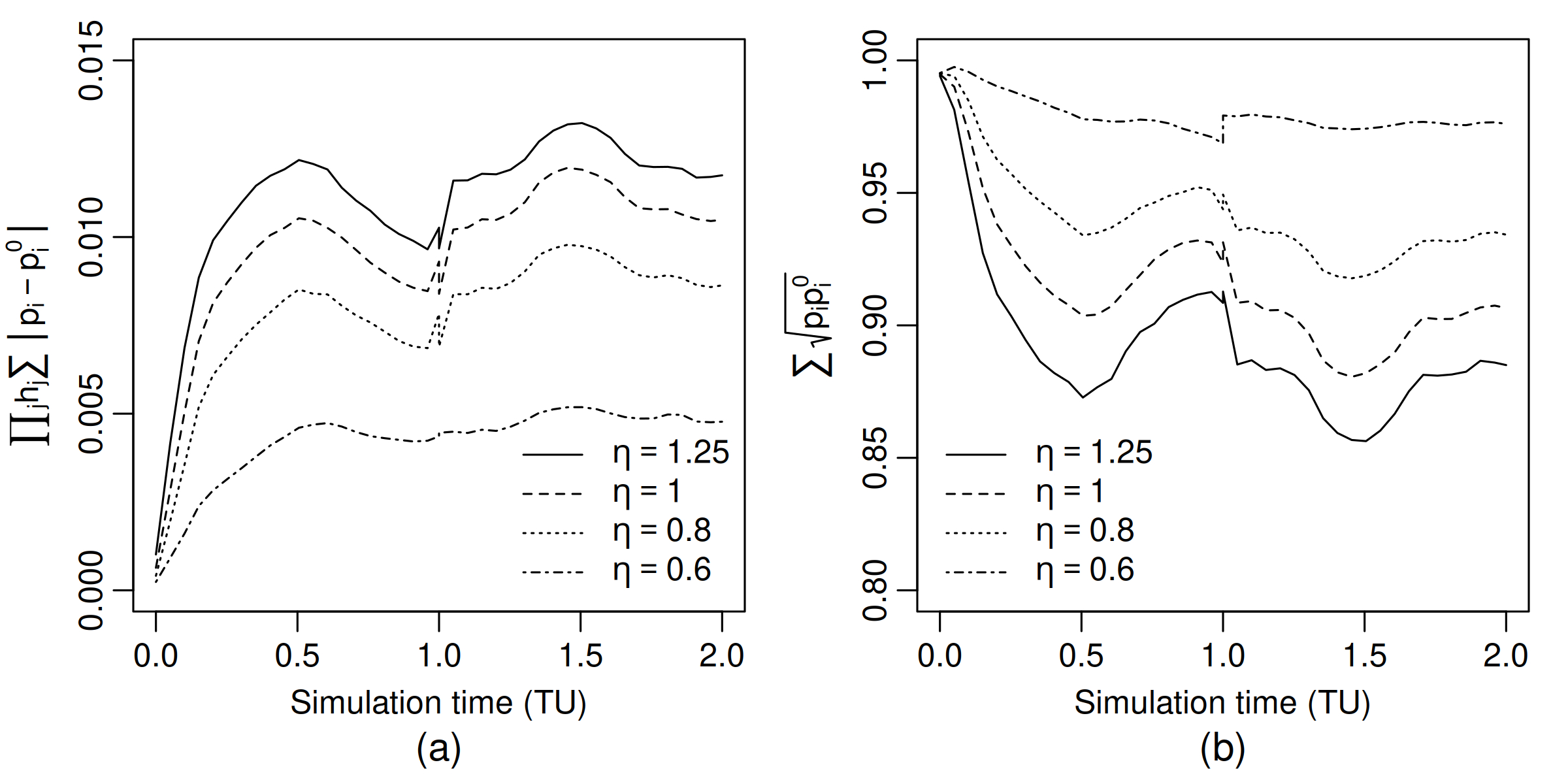}
    \caption{Lorenz '63 comparison of the discrete probability distributions for different cell widths (with respect to a reference propagation with $\eta = 0.5$). On the left (a), the comparison is based on the 1-norm $E(t)$. On the right (b), the comparison utilizes the Bhattacharyya Coefficient ${BC}(t)$.}
    \label{fig_validationCellSize}
\end{figure}

These convergence tests have two primary objectives. First, to ensure the new GBEES implementations qualitatively converge, as demonstrated by decreasing errors relative to the reference propagation when step and cell width sizes are reduced. Second, to assess the impact of the non-deterministic execution order of operations in CUDA. In the CUDA architecture, the execution order of different threads is non-deterministic \citep{nvidia-corporation_cuda_2024}. Due to the non-associativity of floating-point arithmetic caused by the rounding errors, small variations in the computed probability distribution are expected. 

For the temporal convergence test, the reference propagation with $\epsilon = 0.01$ was compared to CPU-legacy, CPU-optimized, and GPU approximate propagations. The CPU-legacy used a fixed step size $\Delta t=0.005$ and the CPU-optimized and GPU versions used adaptive step sizes with $\epsilon = $1.0, 0.5, 0.2, and 0.1. The results of these comparisons are shown in Fig. \ref{fig_validationMetrics} for the Lorenz '63 case. The curves demonstrate convergence as defined by Eq. \eqref{eq_convergence}, with progressively more accurate values as the step size is reduced. Additionally, the differences caused by the non-deterministic execution order in CUDA are minimal compared to the influence of other error sources in the simulation, such as step size variation.

For the spatial convergence test, we run simulations with varying grid cell widths as functions of the default width, or $h_j = \eta \times h_j^*$. As the non-deterministic effects of the GPU version of GBEES were trivial, we use only this version for this test. The reference propagation with $\eta = 0.5$ was compared to propagations with $\eta = 1.25$, $1$, $0.8$, and $0.6$. A CFL number of $C=0.1$ was set in the simulations for this validation. Fig. \ref{fig_validationCellSize} indicates that the convergence behavior, in both metrics, is similar to that observed in the temporal convergence analysis.

\section{Performance} \label{sec_performance}

Performance improvements for both the new CPU and GPU versions are evaluated using the same validation cases described in Section \ref{sec_validation}. Their computational effort is summarized in Table \ref{table_problems}. The significant difference in computational load between the two cases is primarily due to the dimensionality of the models: the Lorenz '63 is formulated over a three-dimensional phase space, while the Lorenz '96 uses a six-dimensional one.

\begin{table}[t]
\centering
\begin{tabular}{l r r}
\hline
 & Lorenz '63 & Lorenz '96 \\
\hline
Maximum grid size &  $\approx 24$k & $\approx 50$M \\
Integration steps & $\approx 960$ & $\approx 1050$ \\
Total cell computations & $\approx 6.5$M & $\approx 30$G \\
\hline
\end{tabular}
\caption{Computational burden for the Lorenz '63 and Lorenz '96 models with a variable step size corresponding to a $\Delta t = $1.0$\times$CFL and a cell size equal to half the standard deviation in each dimension of the first measurement.
}\label{table_problems}
\end{table}

\begin{figure}[t]
\centering
\includegraphics[width=3.0in]{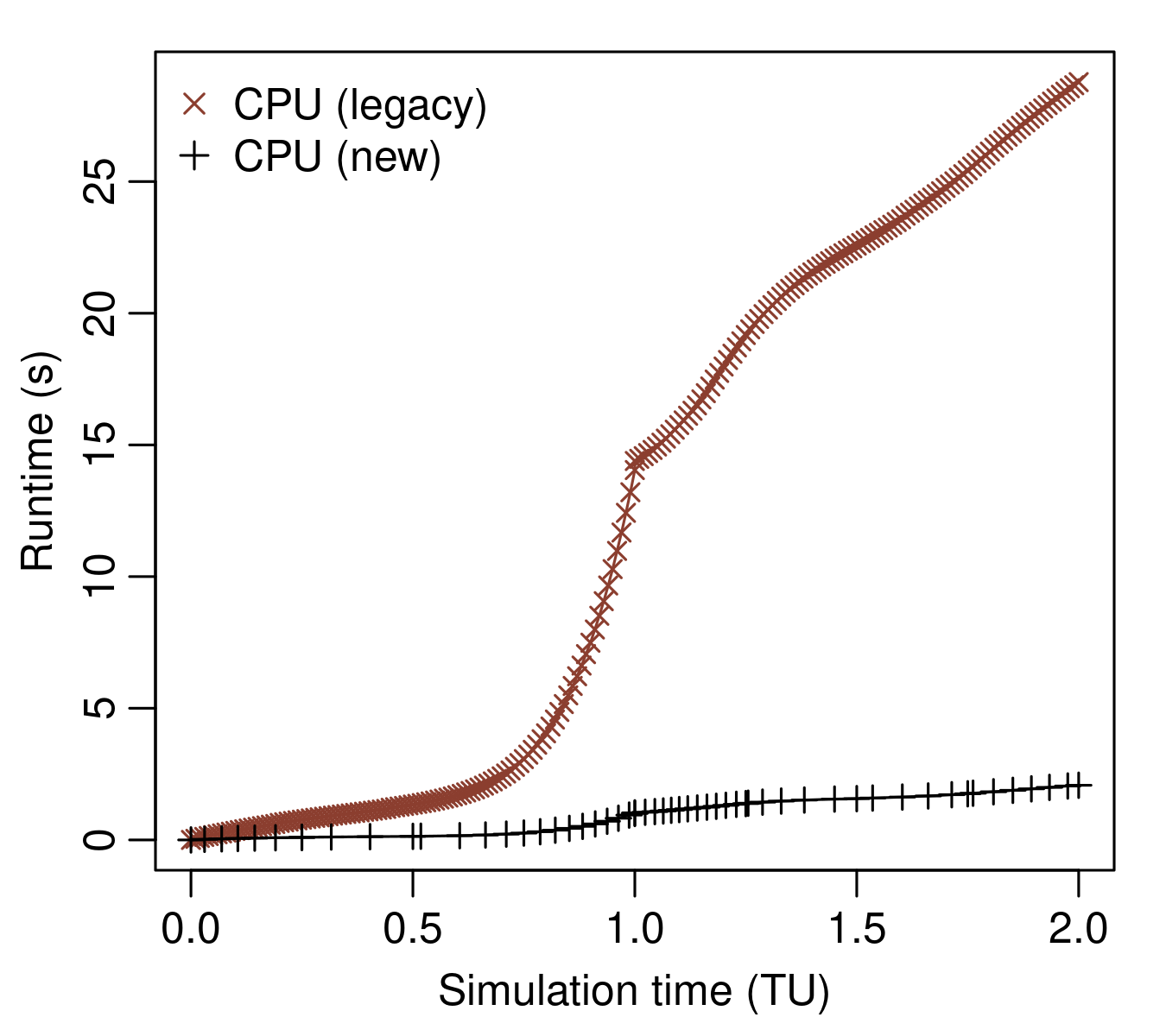}
\caption{Runtime comparison between the legacy and the new CPU version for the Lorenz '63 model.}\label{fig_runtime3DLegacy}
\end{figure}

Fig. \ref{fig_runtime3DLegacy} shows the runtime comparison between the legacy and the new CPU versions for the Lorenz '63 model. The legacy version performs a fixed-step integration, while the new CPU version uses a variable-step scheme. To ensure a fair comparison, the fixed step size of the legacy CPU version was adjusted so that, during the integration, the BCs, computed using the same procedure as in Section \ref{sec_validation}, reach a similar maximum value.

The runtime results of these executions are also included in Table \ref{table_performance_lorenz3D}. The relative speed-up of the new CPU version relative to the legacy version is approximately 13.85 times faster (calculated as $1/0.072$). To assess performance in the CUDA version, it is essential first to outline the launch configuration parameters and explain how these settings influence overall performance. The GPU launch configuration is determined by three key parameters: the number of blocks, the number of threads per block, and the number of cells each thread processes. The objective of the launch configuration is to maximize the GPU occupancy. Since we have a Cooperative Kernel, this is achieved by launching a total number of threads equal to the GPU's maximum simultaneous thread capacity.

\begin{table}[!ht]
\centering
\begin{tabular}{l r r r}
\hline
Device & Runtime (ms) & Cells/s & Speed-up \\
\hline
CPU-legacy: Apple M2 MAX & 28777 & $\approx$0.54M/s & 0.072 \\
CPU-optimized: Apple M2 MAX & 2077 & $\approx$3.13M/s & 1 \\
GPU 1: NVIDIA Tesla V100 & 244 & $\approx$26.6M/s & 8.5 \\
GPU 2: NVIDIA A100 & 258 & $\approx$25.2M/s & 8.1 \\
GPU 3: NVIDIA H100 & 226 & $\approx$28.8M/s & 9.2 \\
GPU 4: NVIDIA H200 & 230 & $\approx$28.3M/s & 9.0 \\
\hline
\end{tabular}
\caption{Total runtime, number of cells processed per second, and relative speed-up compared to a single-core CPU running the optimized version of GBEES for the Lorenz '63 model. See \ref{appendix_hardware} for full specifications of processing units.}
\label{table_performance_lorenz3D}
\end{table}

If the maximum grid size exceeds this capacity, each thread must process multiple cells, requiring the parameter for cells processed per thread to be set to a value greater than one. Conversely, if the maximum grid size is smaller than this capacity, the configuration should launch only as many threads as the grid requires, with each thread processing only one cell. In this last case, the achieved occupancy will be less than the theoretical maximum because the model does not expose sufficient parallelism to fully utilize the GPU. Therefore, the launch configuration strategy is to keep the number of cells processed by each thread as low as possible. If the model is sufficiently large, the product of the number of blocks and the number of threads per block should equal the GPU's maximum thread capacity. This product can be achieved through various combinations of blocks and threads per block.

This balance between blocks and threads per block is very subtle. Setting the number of threads per block to the maximum (1024 in the current CUDA architectures) benefits the parallel reduction and scan processes described in Section \ref{sec_specific}, as more computation is performed at the block level using shared memory. However, the performance differences are minimal, and in some tests, using fewer threads per block than the maximum has resulted in slightly better runtimes.

\begin{figure}[t]
\centering
\includegraphics[width=5.4in]{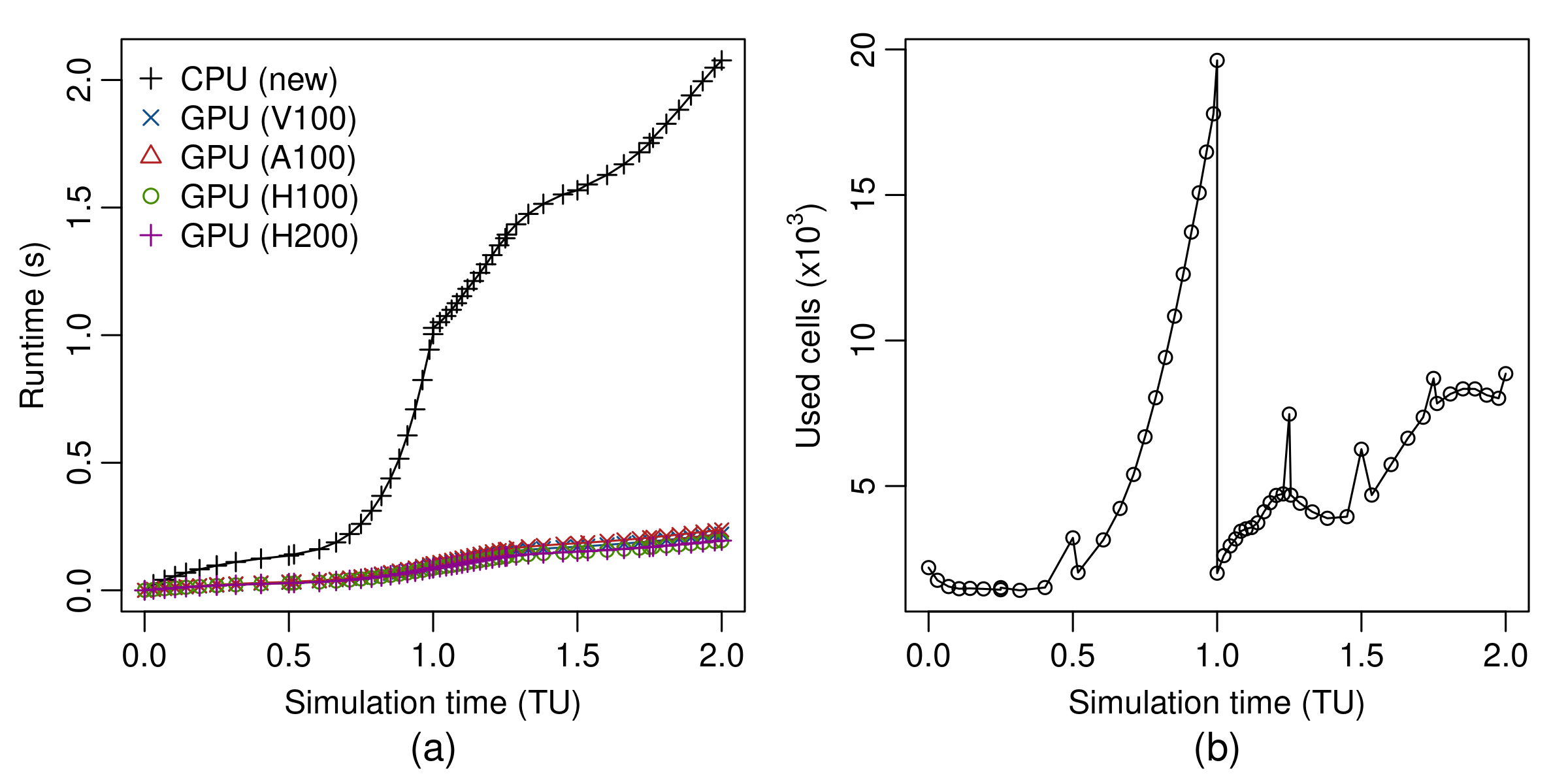}
\caption{Runtime comparison between the new CPU version and GBEES-GPU on various GPU devices (a) and number of used cells (b) for the Lorenz '63 model.}\label{fig_runtime3D}
\end{figure}

Fig. \ref{fig_runtime3D} represents, for the Lorenz '63 model, the runtime comparison between the new CPU version and the CUDA implementation running on different GPU devices. Fig. \ref{fig_runtime3D}(a) represents the program runtime as a function of the simulated time, where steeper regions correspond to moments when the grid contains more cells. The number of cells during the simulation is plotted in Fig. \ref{fig_runtime3D}(b). The drop in the number of cells at $t = 1.0 \text{ TU}$ corresponds to a discrete measurement update.

The graph in Fig. \ref{fig_runtime3D}(a) shows a significant performance boost from parallelizing and executing the algorithm on the GPU. Table \ref{table_performance_lorenz3D} summarizes the total runtime, the number of processed cells per second, and the speed-up values. The Lorenz '63 model represents a case where the grid size is not large enough to achieve maximum occupancy on the tested GPUs. This limitation causes the performance to be similar across all devices. Despite this, the speed-up achieved by using the CUDA version ranges from 8.5 to 9.0, depending on the specific GPU tested.

\begin{figure}[t]
\centering
\includegraphics[width=3.0in]{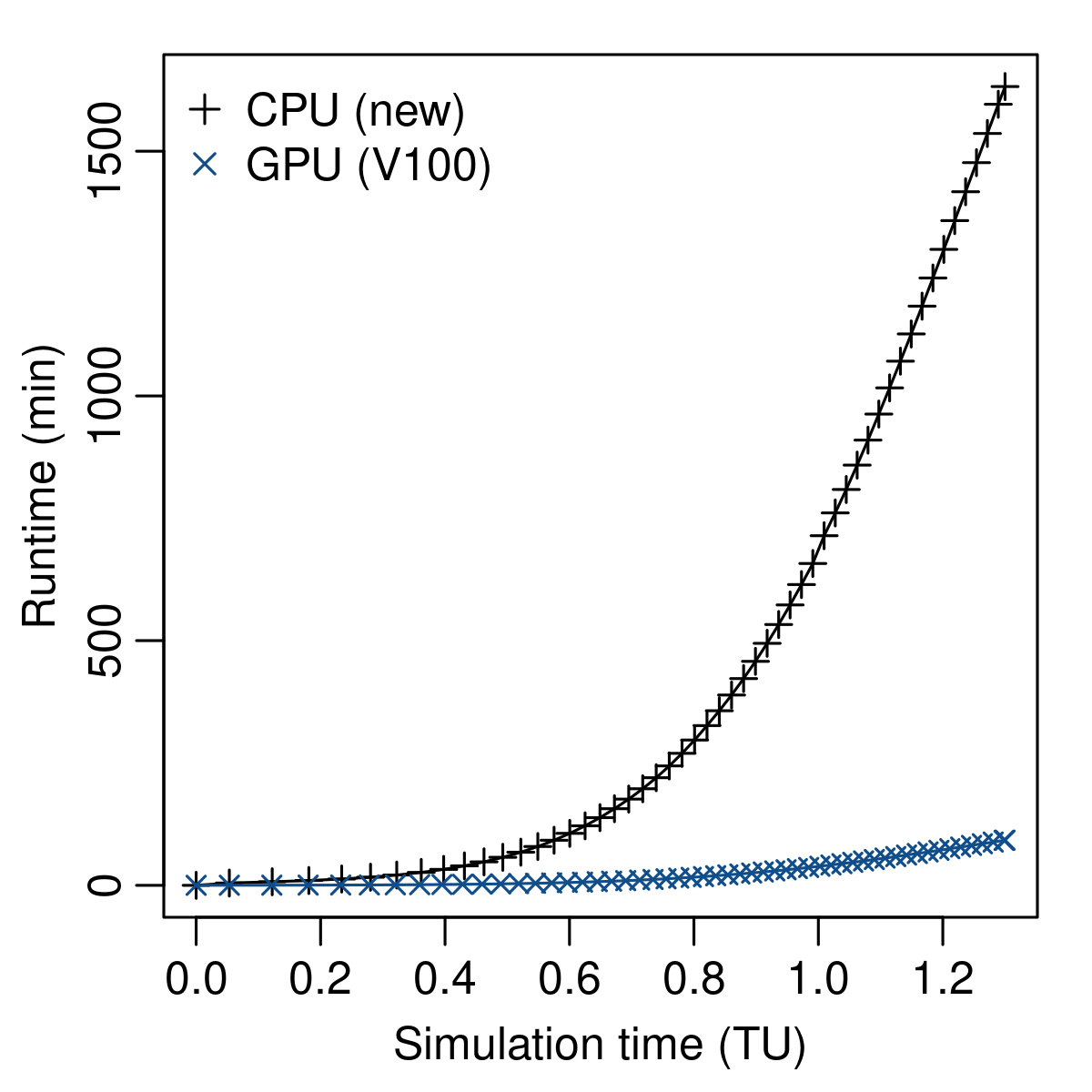}
\caption{Runtime comparison between the new CPU version and the GBEES-GPU execution on a V100 GPU decvice for the Lorenz '96 model.}\label{fig_lorenz6Dcpu}
\end{figure}

The Lorenz '96 model requires a high computational burden and exposes enough parallelism to fully utilize the performance of the tested GPUs. This results in a significant performance difference between the CPU and CUDA versions. To highlight this difference and facilitate representation, Fig. \ref{fig_lorenz6Dcpu} first presents the runtime comparison between the new CPU version and the CUDA implementation running on a V100 GPU device, which is the slowest among the tested GPUs.

\begin{figure}[t]
\centering
\includegraphics[width=5.4in]{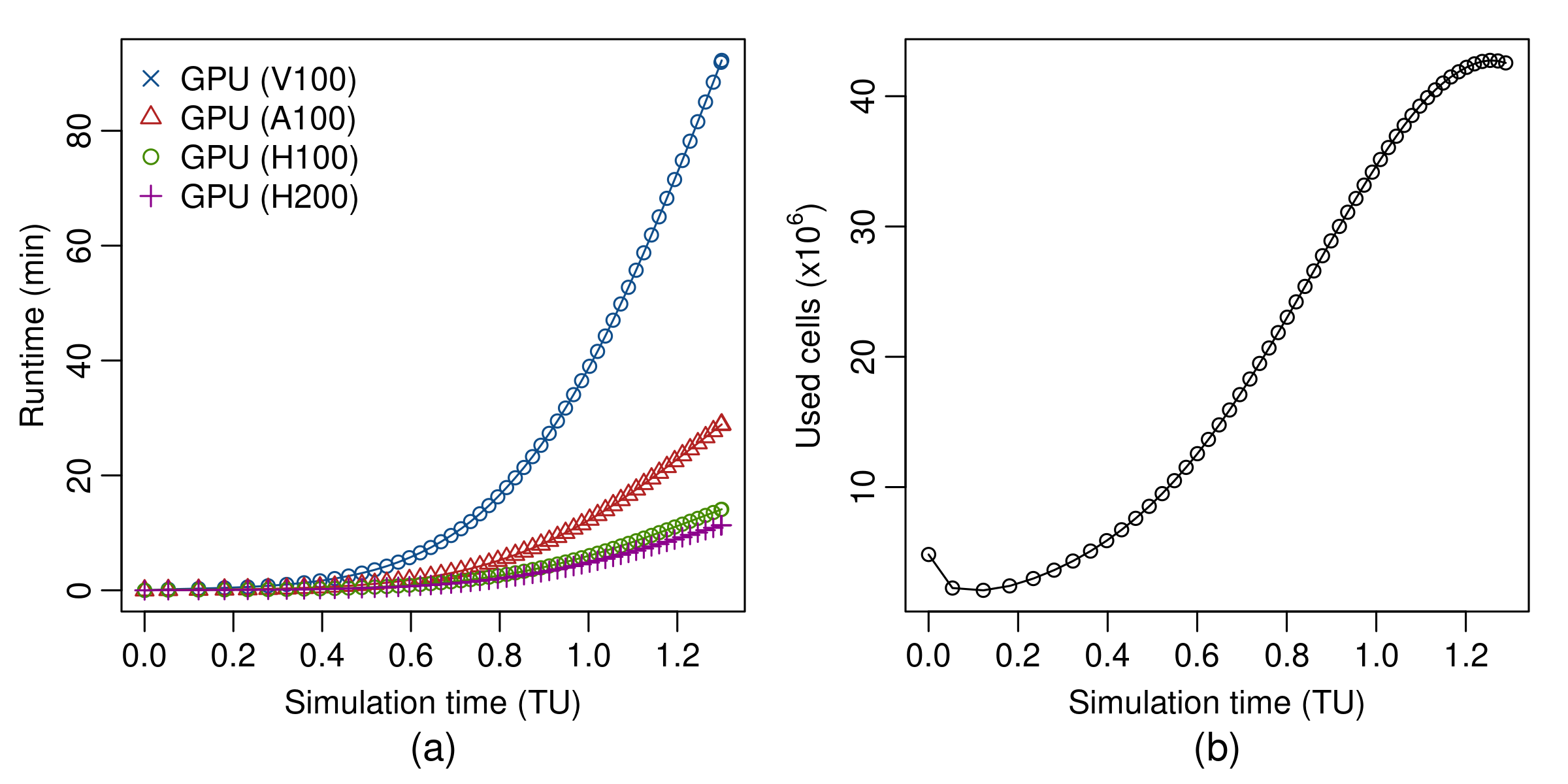}
\caption{Runtime comparison between the new CPU version and GBEES-GPU on various GPU devices (a) and number of used cells (b) for the Lorenz '96 model.}\label{fig_runtime6D}
\end{figure}

The runtime data for this comparison, along with comparisons to other GPU devices, are included in Table \ref{table_performance_lorenz6D}. The execution of the Lorenz '96 model is 17.8 times faster in the CUDA version on the V100 device compared to the new CPU version.
For the other devices, Fig. \ref{fig_runtime6D} shows that, as the model fully utilizes the GPUs, there is a progressive reduction in execution times corresponding to the increasing computing power of the different test devices. The speed-ups achieved are 56.4 times for the A100 device, 106.6 times for the H100, and 132.5 times faster for the H200 device.

\begin{table}[t!]
\centering
\begin{tabular}{l r r r}
\hline
Device & Runtime (s) & Cells/s & Speed-up \\
\hline
CPU-optimized: Apple M2 MAX & 97927 & $\approx$0.3M/s & 1 \\
GPU 1: NVIDIA Tesla V100 & 5513 & $\approx$5.4M/s & 17.8 \\
GPU 2: NVIDIA A100 & 1736 & $\approx$17.3M/s & 56.4 \\
GPU 3: NVIDIA H100 & 919 & $\approx$32.6M/s & 106.6 \\
GPU 4: NVIDIA H200 & 739 & $\approx$40.6M/s & 132.5 \\
\hline
\end{tabular}
\caption{Total runtime, number of cells processed per second, and relative speed-up compared to a single-core CPU running the new version of GBEES for the Lorenz '96 model.}\label{table_performance_lorenz6D}
\end{table}

The observed execution time improvements surpass one order of magnitude between the legacy and new CPU versions ---as demonstrated in the Lorenz '63 use case--- and two orders of magnitude between the new CPU version and the GPU implementation. Together, these results indicate that the enhanced GBEES algorithm achieves a total performance improvement of more than three orders of magnitude.

\section{Conclusions} \label{sec_conclusions}

This paper presents a CPU-optimized implementation and a GPU implementation of GBEES, a second-order accurate, Eulerian algorithm for robust nonlinear uncertainty propagation. To address the computational limitations associated with the legacy CPU implementation of GBEES, the main data structure was changed from a linked list to a hashtable, a CFL-minimized adaptive step size was used, and the grid growing and pruning procedures were adjusted to consider the advection direction. Once the CPU implementation was optimized, the algorithm was translated to CUDA for single GPU execution.

The CUDA implementation is heavily influenced by the dynamic nature of the grid required by the GBEES method. This dynamic grid demands more sophisticated synchronization mechanisms and an efficient memory layout for its storage. To address these challenges, the CUDA implementation employs the Cooperative Kernel abstraction, an ad-hoc data structure to store the grid, non-blocking algorithms to modify it, and parallel-specific optimization techniques.

We validate the two novel implementations using accuracy and convergence. To measure accuracy, we calculate the Bhattacharyya Coefficient (BC) of the GBEES distribution compared with a dense MC distribution, assuming values greater than $0.9$ represent sufficient similarity. To measure convergence, we evolve a truth distribution of GBEES using a time step that is $1/100^{\text{th}}$ of the CFL-based stability limit. We then compare various GBEES implementations, beginning with the CFL time step then decreasing in step size, evaluating convergence through both the 1-norm and the BC. We deem ``convergence'' as the approach of the reference propagation as the adaptive step size is reduced.

The framework is applied to two chaotic systems: the first is the three-dimensional Lorenz '63 model. For this use case, the BC remains above 0.9 for each of the GBEES implementations. The inability of the MC distribution to assimilate corrections emphasizes its inefficacy when a measurement model is present. Additionally, both metrics indicate convergence. The CPU-optimized implementation has a performance increase of $14\times$ relative to the CPU-legacy implementation and the GPU implementation has a performance increase of $9\times$ relative to the CPU-optimized implementation.

Finally, we demonstrate the full capability of the new implementations on a six-dimensional variation of the Lorenz '96 model, an $n$-dimensional chaotic system. For this use case, the BC for the CPU-optimized and GPU implementations compared with a dense MC distribution remains above $0.9$ (implementing this example with the CPU-legacy version is computationally infeasible). Convergence metrics also demonstrated convergence. The high dimensionality of the system highlights the efficacy of the parallelized algorithm; the GPU implementation has a performance increase of $133\times$ relative to the CPU-optimized implementation, implying a 1000-fold increase in performance relative to the CPU-legacy implementation. 

\section*{CRediT authorship contribution statement}

\textbf{Benjamin L. Hanson:} Writing -- original draft, Methodology, Software, Validation, Visualization. \textbf{Carlos Rubio:} Writing -- original draft, Methodology, Software, Validation. \textbf{Adri\'an Garc\'ia-Guti\'errez:} Validation. \textbf{Thomas Bewley:} Writing -- review \& editing, Conceptualization, Methodology, Supervision, Project administration. 

\section*{Declaration of competing interest}

The authors declare that they have no known competing financial interests or personal relationships that could have appeared to influence the work reported in this paper.

\section*{Code availability}

In the interest of facilitating further research, promoting its use, and allowing the reproduction of the experiments, the complete source code is available in the following repositories:
\begin{itemize}
\item The GBEES CPU-optimized code is available at \url{https://github.com/bhanson10/gbees} under the BSD 3-Clause License.

\item The GBEES-GPU code is available at \url{https://github.com/Cx-Rubio/gbees-cuda} under the BSD 3-Clause License.
\end{itemize}

The provided software can be easily extended with new dynamic models. Documentation on how to compile and execute the software, the format of the input and output data, and instructions for extending the software by adding additional models can be found in the repositories themselves and in the corresponding instruction guide \cite{hanson_guide_2025}.

\section*{Acknowledgements}

The authors thank the San Diego Supercomputer Center for a computing time allocation on the Triton Shared Computing Cluster. 

\appendix
\section{Benchmark hardware specifications}
\label{appendix_hardware}

The CPU execution times were measured on a system with the following specifications:
\begin{itemize}
\item CPU: Apple M2 Max, 12-core CPU. Clock frequency 8 cores $\times$ 3.7GHz, 4 cores $\times$ 3.4 GHz. L2 cache size 36 MB. 
\end{itemize}

The GPU performance tests were executed in the next GPU devices:
\begin{itemize}
    \item GPU 1: NVIDIA Tesla V100-SXM2-32GB, CUDA architecture Volta. Stream multiprocessors (SMs) 80. Maximum threads per SM 2048. SM clock frequency 1.530 GHz. Memory clock frequency 0.877 GHz. Memory 32 GB HBM2.
    
    \item GPU 2: NVIDIA A100-SXM4-40GB, CUDA architecture Ampere. Stream multiprocessors (SMs) 108. Maximum threads per SM 2048. SM clock frequency 1.410 GHz. Memory clock frequency 1.215 GHz. Memory 40 GB HBM2e.
    
    \item GPU 3: NVIDIA H100-80GB, CUDA architecture Hopper. Stream multiprocessors (SMs) 132. Maximum threads per SM 2048. SM clock frequency 1.980 GHz. Memory clock frequency 2.619 GHz. Memory 80 GB HBM3.
    
    \item GPU 4: NVIDIA H200-141GB, CUDA architecture Hopper. Stream multiprocessors (SMs) 132. Maximum threads per SM 2048. SM clock frequency 1.980 GHz. Memory clock frequency 3.201 GHz. Memory 141 GB HBM3e.
\end{itemize}

\bibliographystyle{elsarticle-num} 
\bibliography{gbees-cuda-rev3}

\end{document}